\documentclass[aps,prd,nofootinbib,superscriptaddress]{revtex4}

\usepackage{epsfig}
\usepackage{multirow}

\newcommand{\be}{\begin{eqnarray}}
\newcommand{\ee}{\end{eqnarray}}

\def\nue{{\nu_e}}
\def\anue{{\bar\nu_e}}

\newcommand{\chr}{\mbox{$\breve{\rm C}$erenkov~}}

\newcommand{\ms}{\Delta m^2_{21}}
\newcommand{\ma}{\Delta m^2_{31}}

\newcommand{\sss}{\sin^2 \theta_{12}}
\newcommand{\sch}{\sin^2 \theta_{13}}
\newcommand{\stch}{\sin^2 2\theta_{13}}
\newcommand{\sa}{\sin^2 \theta_{23}}

\newcommand{\bb}{$\beta$-Beam~}

\newcommand{\br}{$^8$B~}
\newcommand{\li}{$^8$Li~}
\newcommand{\he}{$^6$He~}
\newcommand{\neon}{$^{18}$Ne~}
\def\ltap{\ \raisebox{-.4ex}{\rlap{$\sim$}} \raisebox{.4ex}{$<$}\ }
\def\gtap{\ \raisebox{-.4ex}{\rlap{$\sim$}} \raisebox{.4ex}{$>$}\ }
\newcommand{\ie}{{\it i.e.}}

\newcommand{\deltacp}{\delta_{\mathrm{CP}}}

\begin{document}
%%%%%%%%%%%%%%%%%%%

\begin{flushright}
    %{\tt hep-ph/ }\\
    {IFT-UAM/CSIC-09-33} \\
    {MPP-2009-118}\\
    {EUROnu-WP6-09-09}\\
    {CUP-TH-09-01}
\end{flushright}

\title{Optimized Two-Baseline Beta-Beam Experiment}

\author{Sandhya Choubey}
\email[]{sandhya@hri.res.in}
\affiliation{Harish-Chandra Research Institute, Chhatnag Road, Jhunsi, Allahabad  211019, India}

\author{Pilar Coloma}
\email[]{p.coloma@uam.es}
\affiliation{Departamento F\'{\i}sica Te\'orica and Instituto F\'{\i}sica Te\'orica UAM/CSIC, Cantoblanco, E-28049 Madrid, Spain}

\author{Andrea Donini}
\email[]{andrea.donini@uam.es}
\affiliation{Instituto F\'{\i}sica Te\'orica UAM/CSIC, Cantoblanco, E-28049 Madrid, Spain}

\author{Enrique Fernandez-Martinez}
\email[]{enfmarti@mppmu.mpg.de}
\affiliation{Max-Planck-Institut f\"ur Physik (Werner-Heisenberg-Institut), F\"ohringer Ring 6, 80805 M\"unchen, Germany}

\begin{abstract}
\vglue 0.9cm
We propose a realistic $\beta$-Beam experiment with four source ions 
and two baselines for the best possible sensitivity to $\theta_{13}$, 
CP violation and mass hierarchy. Neutrinos from \neon and 
\he with Lorentz boost $\gamma=350$ are detected in a 500 kton water \chr 
detector at a distance $L=650$ km (first oscillation peak) from the source. Neutrinos from 
\br and \li are detected 
in a 50 kton magnetized iron detector at a distance $L=7000$ km 
(magic baseline) from the source. Since the decay ring requires a tilt 
angle $\vartheta=34.5^\circ$ to send the beam to the magic baseline, 
the far end of the ring has a maximum depth of $d=2132$ m for magnetic 
field strength of 8.3 T, if one demands that the fraction of ions that
decay along the straight sections of the racetrack geometry 
decay ring (called livetime) is 0.3. We alleviate this 
problem by proposing to trade reduction of the livetime 
of the decay ring with the 
increase in the boost factor of the ions, such that the number of events at 
the detector remains almost the same. This allows to substantially reduce the 
maximum depth of the decay ring at the far end, without significantly compromising 
the sensitivity of the experiment to the oscillation parameters. We take 
\br and \li with $\gamma=390$ and 656 respectively, as these are the largest 
possible boost factors possible with the envisaged upgrades of the SPS at
CERN. This allows us to reduce $d$ of the decay ring by a factor of 1.7 for
8.3 T magnetic field. Increase of magnetic field to 15 T would further 
reduce $d$ to 738 m only. We study the sensitivity reach of this two baseline 
two storage ring $\beta$-Beam experiment, and compare it with the
corresponding reach of the other proposed facilities. We find that for values of $\sin^2 2 \theta_{13} > 10^{-3}$ this $\beta$-Beam setup outperforms the Neutrino Factory sensitivities.
\end{abstract}
\pacs{}
\preprint{}
\maketitle
%\newpage
%%%%%%%%%%%%%%%%%%%%%%%%%
\section{Introduction}
%%%%%%%%%%%%%%%%%%%%%
Neutrinos have been providing some of the 
most illuminating as well as intriguing insights into the 
theory of elementary particle physics. 
Neutrinos are 
naturally massless within the framework of the Standard 
Model of particle physics. The presence of tiny neutrino masses 
therefore demands for a theory beyond the Standard Model. 
The neutrino mass and mixing pattern, once determined to 
a sufficient level of accuracy could (hopefully) show the path to this 
theory underlying the Standard Model. 
Existence of neutrino masses and mixings has now been established
by a series of outstanding experimental 
endeavors involving neutrinos coming from the Sun \cite{solar},
Earth's atmosphere \cite{atm}, nuclear reactors 
\cite{kl} and accelerator sources \cite{k2k,minos}. The 
global neutrino data prefers \cite{limits} 
$\ms = 7.7\times 10^{-5}$ eV$^2$,
$\ma = 2.4\times 10^{-3}$ eV$^2$, $\sss = 0.31$ and 
$\sa = 0.5$. 

The third mixing angle $\theta_{13}$ is mainly
constrained by the results from the Chooz reactor 
experiment \cite{chooz}, which is consistent with no 
positive signal for oscillations and hence a zero 
value for this mixing angle. This data, when combined 
with the world neutrino data, gives $\sch < 0.05$ at the 
$3\sigma$ C.L. However,  
while the Chooz data do not support any evidence 
for non-zero $\theta_{13}$, 
it has been observed that inconsistency between the 
global solar data and KamLAND results can be 
reduced with a non-zero $\theta_{13}$ \cite{th13hint} 
(see also \cite{th13hint2}). While the evidence is still 
weak\footnote{There also might be an indication of non-zero 
$\theta_{13}$ coming from the Super-Kamiokande atmospheric neutrino 
data, however this claim is still considered to be controversial 
(see \cite{th13hint} for a detailed discussion).}, 
we do have an indication for non-zero $\theta_{13}$ 
at the $1\sigma$ level. This claim for non-zero $\theta_{13}$ 
could be {\it just} within the reach of the next-generation 
neutrino experiments involving reactor antineutrinos,  
\cite{chooz2,dayabay,reno,angra} 
and accelerator (anti)neutrinos \cite{t2k,nova}. In Fig. 1 of 
Ref.~\cite{Huber:2009cw} the $90 \%$ CL. sensitivities to $\theta_{13}$, CP violation
and the mass hierarchy expected for these next generation of experiments are presented 
as well as the sensitivities achievable by an eventual combination of all of them. The combination of all the
facilities would grant sensitivity to $\theta_{13}$ down to $\sin^2 2 \theta_{13} > 3-6 \times 10^{-3}$ allowing to probe 
the present hint for non-zero $\theta_{13}$. 
If next-generation reactor- and accelerator-based 
experiments fail to observe any positive signal for non-zero 
$\theta_{13}$, however, more powerful experiments 
involving bigger detectors and improved beams in order to pin 
down this elusive mixing angle will be needed. 
Two other oscillation parameters, 
indispensable for the reconstruction of the full neutrino mass 
matrix, are the 
ordering of the neutrino mass ($sgn(\ma)$), {\it aka}  
the neutrino mass hierarchy, and\footnote{The absolute neutrino mass scale and two additional phases (if neutrinos are 
Majorana fermions) are also unknown and required in order to complete the neutrino 
mass matrix. These parameters are inaccessible in 
neutrino oscillation experiments and must be measured 
elsewhere.}  the CP phase $\delta$. It is quite unlikely that 
these will be discovered in the next generation 
experiments. Indeed, as can also be seen from Fig. 1 of 
Ref.~\cite{Huber:2009cw}, the sensitivities to these parameters of the next generation of facilities is very limited even after combining all of them.
CP violation might only be discovered for less than $20\%$ of the possible values of $\delta$ and only if $\sin^2 2 \theta_{13} > 0.02$. As for the mass
hierarchy, it can only be distinguished for less than $40\%$ of the possible values of $\delta$ and only if $\sin^2 2 \theta_{13} > 0.04$. Moreover, these limits are only for a $90 \%$ CL. significance and, therefore, even if the combination of all experiments hints at its precesnce, an independent confirmation at higher significance would be desirable. Therefore, there are good reasons to consider 
larger dedicated experiments, with very well known beams, 
higher statistics and lower systematics and beam backgrounds, even if a signal for non-zero $\theta_{13}$ is found by next generation facilities.

Two kind of such experiments, to improve over the next generation of facilities, have 
been envisaged. The first category, called ``Neutrino Factory"~\cite{geer}, would exploit 
very high intensity neutrino beams coming from the decay of 
muons, which are collected, accelerated and subsequently stored 
in a decay ring. As it is well known, such an experiment necessarily requires 
a far detector with charge identification capability to 
tag the initial neutrino flavor. 
The second kind of high intensity beam proposed is the so-called 
``$\beta$-Beam'' \cite{zucc}. 
This entails producing $\beta$-unstable radioactive 
ions, collecting, bunching, accelerating and then 
finally transferring them into a storage ring \cite{lindroos,betabeampage}. 
In this case, in principle, one could use any 
kind of detector\footnote{The optimum 
choice of the detector depends on the beam characteristics, which we 
will discuss in the next section.}  
with good muon identification. 

For both beam categories, 
the oscillation channel which is expected to give us information on 
all the three parameters is the $\nue \rightarrow \nu_\mu$ channel 
(proportional to the oscillation probability $P_{e\mu}$) -- also called the ``golden channel'' \cite{golden}. 
However, given that only one channel is used to determine 
three parameters and 
that the octant of the atmospheric mixing angle remains 
unknown\footnote{Current experiments 
measure $\sin^2 2\theta_{23}$, and hence two values of 
$\theta_{23}$ can fit the data.}, 
``parameter degeneracies'', which limit the 
sensitivity of the experiment, appear. 
For every true 
set of $\theta_{13}-\delta$ 
the analysis of the data could give 
degenerate solutions arising due to 
(i) the $\delta - \theta_{13}$ correlation  (the so-called ``intrinsic degeneracy''  \cite{intrinsic}),
(ii) the $\delta - sgn(\ma)$ correlation  (the ``sign degeneracy''  \cite{minadeg}), and 
(iii) the octant of $\theta_{23}$ (the ``octant degeneracy''  \cite{th23octant}), leading, in total,
to an eight-fold degeneracy in the  $\delta - \theta_{13}$ plane \cite{eight} (the fourth degeneracy being 
the ``mixed degeneracy'', \ie  \, a wrong choice of both $sgn(\ma)$ and of the $\theta_{23}$-octant). 
Notice that the intrinsic and sign (and, hence, the mixed) degeneracies involve the CP phase.
These degeneracies could severely limit the sensitivity to $\delta$ and to other observables
of the experiment, threatening to wash-out all the advantages coming from the well known, 
high intensity beam source. 
Extensive efforts have been made to constrain the fake solutions 
and thereby improve the sensitivity of these expensive 
experiments. A variety of ways have been suggested in the literature.
These include combining experiments using the golden channel 
but at different baselines \cite{diffLnE, t2ksimulation}, 
combining different channels 
\cite{silver,dissappear,pee} at long baseline experiments, and 
combining data from other type of experiments 
\cite{addatm,cernmemphys,addreact}. 
It has been noted \cite{magic,magic2,petcov} that $P_{e\mu}$ can be made 
independent of $\delta$ at a baseline which corresponds to the
characteristic oscillation wavelength of the 
neutrinos in Earth matter, the so-called ``magic baseline'' \cite{magic}.
This baseline can be shown to be independent of all oscillation 
parameters, as well as the neutrino energy. Since the $P_{e\mu}$ 
probability at this baseline is independent 
of $\delta$, and hence is not affected by the 
intrinsic, sign and mixed degeneracies, 
it provides a clean bedrock for the measurement of $\theta_{13}$ 
and $sgn(\ma)$. In fact, 
an optimizing exercise in $L$ confirms that a magic baseline 
experiment is indeed the best baseline option to measure
$\theta_{13}$ and $sgn(\ma)$ for both Neutrino Factory and 
high energy $\beta$-Beam. However, since CP violation cannot be measured at the 
magic baseline, 
we should perform the experiment at an additional baseline 
which has good sensitivity to $\delta$. Optimization 
studies have revealed that this baseline turns out to be 
about 4000 km for the Neutrino Factory \cite{nufactoptim}. For the 
$\beta$-Beam, the  choice of this second baseline is more involved and 
depends on the choice of the $\beta$-unstable ion at the 
source and on the boost factor $\gamma$ \cite{bboptim}. 

Two sets of radioactive ions have been considered extensively in the literature
as possible $\beta$-Beam sources: \neon (for $\nue$)  
and \he (for $\anue$), which were introduced in the pioneering 
$\beta$-Beam beam proposal by Piero Zucchelli \cite{zucc}; 
and  \br (for $\nue$) and \li (for $\anue$) 
\cite{rubbia,mori}. The main difference between the two sets 
lies in their Q-values, which is about 4 times larger for the latter set. 
Therefore, neutrino beams produced through the decay of \br and \li
have an energy about 4 times larger than those produced with the decay
of \neon and $^6$He, when using the same boosting factor. 
%({\it i.e.} the same facilities to accelerate the ions). 
It was shown in \cite{betaino1,betaino2} that 
using \li and \br ions and performing the experiment at the 
magic baseline returns excellent sensitivity to $\theta_{13}$ and 
the mass hierarchy. A baseline optimization study \cite{bboptim} 
showed that indeed the magic baseline is the best 
place to measure these two oscillation parameters if 
one uses a $\beta$-beam fueled with \br and \li ions. 
The mass hierarchy, in particular, demands a $\beta$-beam set-up 
with \br and \li as the source ions and a detector located 
at the magic baseline.  
This is easily explained: for the boost factor $\gamma \sim 350$, the neutrinos
produced in the $\beta$-decay of these ions are seen to 
have energies peaked in the range $E_\nu \in [4,6]$ GeV, where 
$P_{e\mu}(E_\nu)$ picks up near-resonant matter effects and 
becomes very large. On the other hand, neutrinos produced 
in the $\beta$-decay of \he 
and \neon ions with the same $\gamma$ are peaked at only 
$E_\nu \sim 1.5$ GeV, and $P_{e\mu}$ is non-resonant. 
As it was the case for Neutrino Factory beams, since the magic 
baseline smothers the CP dependence 
of $P_{e\mu}$, the CP violation studies will require another baseline. 
It was seen \cite{bboptim} that $L \simeq 1000 - 2000$ km  
was the optimal choice for CP studies if one uses \br and 
\li also for this second baseline. On the other hand, if one uses 
the lower Q \neon and \he ions, the best results come at 
$L\simeq 600 - 700$ km. Sensitivity reach for a two-baseline 
$\beta$-Beam set-up with \br and \li as source ions 
and 50 kton magnetized iron detector at both 
baselines ($L=7000$ km and $L=2000$ km) was studied in 
\cite{twobaseline1}. Another two-baseline set-up, using \br and \li 
as source for a 50 kton magnetized iron detector at the magic baseline 
and \neon and \he as the source for a 50 kton Totally Active Scintillator 
Detector (TASD) at $L=730$ km was proposed in \cite{twobaseline2}. 
The sensitivity reach for both two-baseline \bb set-ups was 
seen to be remarkable, and for very high values 
for the number of useful radioactive ion decays and $\gamma$, 
even comparable to the Neutrino Factory. 

However, all studies on the \bb set-ups with the magic baseline 
as the source-detector distance suffer from one major drawback. 
They require $E_\nu$ to be peaked around [4-7] GeV. In order to 
produce such high neutrino beams, even \br and \li will have to be 
accelerated to $\gamma \gtap 350$. It is seen 
that the sensitivity of the experiment depends crucially 
on high boost factors, and falls sharply as  
$\gamma$ drops below 350 \cite{betaino2}. 
Such high values of $\gamma$ not only demand bigger accelerators, 
they make the bending of the source ions difficult in the storage ring. 
While the acceleration of the ions could be feasible with the planned 
replacement of the SPS at CERN with a new machine, the SPS+  \cite{bc2,PAF}, 
the demand on the storage ring appears to be rather unrealistic. 
The original storage ring design \cite{zucc}, for \he and \neon 
boosted at $\gamma = 100$, consisted of a racetrack ring with straight
section of $L_s = 2500$ m, curved sections of radius $R = 300$ m 
(using 5 T magnets) and a total ring length $L_r = 6885$ m. 
As it has been be discussed before \cite{twobaseline1}, 
and will be again discussed at length in this paper, 
$\gamma = 350$ for \he ($\gamma=390$ for $^8$Li) requires 
a racetrack storage ring with curved sections of radius $R\sim 632$ m, 
exploiting LHC magnets with maximum magnetic field of 8.3 T. 
If we keep the straight sections of the racetrack storage ring unchanged,
we end up with a ring with a longitudinal section of 3764 m, 
to be compared with the 3100 m of the original design. 
The main problem in using this ring design to send a  \li beam to  
$L=7000$ km is the following: the ring plane should be tilted by $34.5^\circ$ 
inside the Earth for the beam to be shot at the magic baseline, and the corresponding
depth of the ring at its far end is a whopping $d=2132$ m. 
This is well beyond any realistic possibility. 

In this paper, we modify the two-baseline \bb set-up in order to alleviate this problem with the storage 
ring. We propose a more realistic \bb set-up for the next to next generation of neutrino oscillation facilities, where we produce, accelerate and store ions of the 
four kinds (\he, \neon, \li and \br) at CERN, each of them running for a period of 2.5 years. 
The experiment time-length is therefore of 10 years in total.
We aim \he- and \neon-generated low-energy neutrino beams to a megaton water \chr detector located at $L=650$ km
from the source, possibly at Canfranc in Spain, and \li- and \br-generated high-energy neutrino beams to a 50 kton 
iron detector at a distance close to the magic baseline, possibly at INO in India.  
We use two separate storage rings for this purpose. The first ring corresponds to the design sketched above, 
called hereafter ``the long ring", with $L_s = 2500$ m straight sections and $R = 632$ curvature radius and 
it will be used to store \he and \neon ions boosted at $\gamma = 350$. 
The same ring design, with 8.3 T magnets, can be used to store \li (\br) ions accelerated up to $\gamma =390$ 
($\gamma = 656$). For \li ions such a small increase in the boost factor corresponds, on the other hand, to a 50\% increase 
of the statistics that can be collected in the far detector. We can, therefore, design a dedicated ring to aim at the far detector with smaller straight sections, such that the maximal depth $d$ of the far end of the ring can be made smaller than 
for the ``long ring". The price to pay is that the total number of useful \br and \li decays towards the 
iron detector at INO is reduced by 40\%. We show that this does not drastically impair the performance of the experiment, 
and that actually the sensitivity to the mass hierarchy is similar to  that that can be obtained  with the ``long ring" with 
\li and \br boosted at $\gamma = 350$. This ring will be called hereafter ``the short ring". 

Since detailed results on the response of the iron detector for this kind of 
experiment are unavailable, we assume very conservative estimates for the energy threshold, 
energy resolution and backgrounds. We study the effect of these detector characteristics on the sensitivity reach 
of the experiment. 

The paper is organized as follows. In section II, we 
discuss in more detail the experimental set-up, and in 
particular propose the modified ``short" storage ring for \br and \li aimed at the magic baseline.  
In section III we present our results and compare the sensitivity 
of our modified two-baseline set-up against some of the 
other high $\gamma$ \bb options proposed and studied before.   
In section IV we study the effect of reducing the 
storage ring length, by comparing 
our results with one ``long" and one ``short" ring with
those where two identical ``long" storage rings are assumed. 
We also study the effect of ``improved'' iron detector characteristics.  
Finally, we present our conclusions in section V. 

%%%%%%%%%%%%%%%%%%%%%%%%%%%%%%%%%%%%%%%%%%%%%%%%%
\section{Two-Baseline $\beta$-Beam Experiment}
%%%%%%%%%%%%%%%%%%%%%%%%%%%%%%%%

In this section we will discuss in more detail the 
various aspects related to the $\beta$-Beam experiment. 
As stated before in 
the Introduction, we have two widely accepted set of candidate 
source ions which could be effectively used to produce a high 
intensity beam. We give the characteristics of these ions in 
Table \ref{tab:ions}. 
\begin{table}
\begin{center}
\begin{tabular}{|c|c|c|c|c|} \hline \hline
   Element  & $A/Z$ & $T_{1/2}$ (s) & $E_0$ eff (MeV) & Decay Fraction \\ 
\hline
  $^{18}$Ne &   1.8 &     1.67      &        3.41         &      92.1\%    \\
            &       &               &        2.37         &       7.7\%    \\
            &       &               &        1.71         &       0.2\%    \\ 
%  $^{19}$Ne &   1.9 &    17.2       &        2.31         &       100\%    \\
  $^{8}$B   &   1.6 &     0.77      &       13.92         &       100\%    \\
\hline
 $^{6}$He   &   3.0 &     0.81      &        3.51         &       100\%    \\ 
 $^{8}$Li   &   2.7 &     0.83      &       12.96         &       100\%    \\ 
\hline
\hline
\end{tabular}
\caption{\label{tab:ions} $A/Z$, half-life and end-point energies for three $\beta^+$-emitters ($^{18}$Ne  and $^8$B)
and two $\beta^-$-emitters ($^6$He and $^8$Li). All different $\beta$-decay channels for $^{18}$Ne are presented~\cite{beta}.}
\end{center}
\end{table}
The other aspects which determine the 
(anti)neutrino beam are the number of useful ion decays $N_\beta$  and 
the Lorentz boost $\gamma$. 
It is well known (see for instance \cite{bboptim} for a discussion) 
that for two different isotopes producing a $\nue$ beam, 
if one demands the same spectral shape of the neutrino 
flux, {\it i.e.} the same peak energy and 
normalization, then the following relations hold:
\be
\frac{N_\beta^{(1)}}{N_\beta^{(2)}} \simeq \left( \frac{E_0^{(1)}}
{E_0^{(2)}}\right)^2 \, , \quad 
\frac{\gamma^{(1)}}{\gamma^{(2)}} \simeq \frac{E_0^{(2)}}{E_0^{(1)}} 
\, \qquad  \Rightarrow \qquad
\frac{N_\beta^{(1)}}{N_\beta^{(2)}} \simeq \left( \frac{\gamma^{(2)}}
{\gamma^{(1)}}\right)^2 \, , 
\label{eq:cond}
\ee
where $E_0^{(i)}$ is the end-point energy of the ion-decay, and where we have neglected the 
effect of the electron mass. 
Clearly, the higher the end-point energy of the 
$\beta$-decay of an ion, the lower the $\gamma$ needed
to reach a given neutrino energy in the lab frame. 
Recall that the maximum energy of the neutrino in the 
lab frame is given by $E_\nu^{max} = 2 \gamma (E_0-m_e)$, where 
$m_e$ is the electron mass. 
Therefore, it is easier to reach higher neutrino 
energies using ions with higher end-point energy. 
At the same time, however, to have the same number 
of events in the far detector for two sets of ions
with different $E_0$ boosted at the same $\gamma$, 
we need larger number of useful ion decays
for source ions with higher $E_0$. 
For our candidate source ions we can see 
that the following conditions hold
\be
N_\beta^{B+Li} \simeq 12 \cdot
N_\beta^{Ne+He} \, , \quad
\gamma^{Ne+He} \simeq 3.5 \cdot \gamma^{B+Li}
,
\label{eq:condnum}
\ee 
in order to obtain the same neutrino flux spectrum. 

Experimental challenges on both $N_\beta$ and $\gamma$ are in fact 
intimately related to a large extent. The boost directly depends on the 
amount of acceleration possible. The number of useful ion decays, 
on the other hand, is affected due to losses during the acceleration 
process and hence impacts the amount of acceleration possible. 
Another important way $N_\beta$ and $\gamma$ 
get related is through the design of 
the storage ring. Higher boost factors of the source ions 
make them harder to bend. Thus, for the same magnetic field 
strength, a larger curved section of the storage ring is required to bend ions boosted at high $\gamma$ 
than at low $\gamma$.  Unless the straight sections are increased proportionally, 
the fraction of stored ions that decays in the straight sections of the 
ring (the so-called ``livetime'' {\em l} $=L_s/L_r$) decreases. We will discuss this in detail in section II.B. 

On the other hand, one of the most challenging constraints on the achievable neutrino fluxes comes from the 
requirement of reducing the atmospheric neutrino background in comparison 
to the $\beta$-Beam signal.  The reason is the following: 
in the original $\beta$-Beam proposal, the typical neutrino energy for
neutrinos produced by the decay of $^6$He and $^{18}$Ne ions boosted at
$\gamma = 100$ is $E_\nu \sim 200$ MeV. The number of muons produced by
atmospheric  neutrinos crossing the detector aligned with the $\beta$-Beam flux in this
range of energies was found to be of the order of tens of events per kton per year. 
This background would completely dominate over the oscillation signal.  
Reduction of the atmospheric neutrino background demands 
stringent bunching of the source ions in the decay ring 
so as to pulse the signal in the detector to the required level. 
In order to have a good time correlation of the signal with the neutrino flux produced at the source, 
the ions circulating in the storage ring must occupy a small fraction of the latter. 
The fraction of the ring filled by ions at a given time, also called ``suppression factor" $S_f$, is:  
\begin{equation}
S_f = \frac{ v \times \Delta t_b \times N_b}{L_r} \,
\end{equation}
where $v \sim c$ is the ion velocity, $\Delta t_b$ is the time length of the
ion bunch (the product $v \times \Delta t_b$ is the spatial length of a bunch in the lab frame), 
$N_b$ is the number of circulating bunches  and $L_r$ is the total length of the ring. For
$^6$He/$^{18}$Ne ions boosted at $\gamma = 100$, the suppression factor must be $S_f \sim 10^{-3}$. 
%For $^6$He and $^{18}$Ne accelerated to $\gamma=100$
%the suppression factor needs to be at least $10^{-3}$ 
%in order for the background not to dominate the signal. 
Such a tight $S_f$ can be achieved with a challenging $\Delta t_b = 10$ ns time-length, 
with a maximum of $N_b = 8$ bunches circulating at the same time.
Since both the time-length of the bunch and the number of bunches that can be injected into the ring 
at the same time depends on the details of the acceleration chain and cannot be modified easily, 
a large value of $L_r$ permits to keep $S_f$ at the desired level at the cost  of a bigger ring.
This means that, in turn, only a small $10^{-3}$ fraction of the storage ring is occupied by the ion beam.
Notice that the atmospheric neutrino flux decreases rapidly with energy. In fact, the atmospheric neutrino events are 
known to fall faster than $E_\nu^{-2}$, where $E_\nu$ is the neutrino energy. 
Therefore, for $\gamma=350$ with the same ions, neutrino energies achievable are a factor of 3.5 higher and  
the atmospheric neutrino background reduces by a factor of more than ten. Hence the 
suppression factor needed to smother the atmospheric neutrino background can be relaxed by about an 
order of magnitude to $10^{-2}$. This allows a larger fraction of the storage ring to be used
by the neutrino beam, and $N_\beta$ could be increased consequently. 
In the case of neutrinos from high $\gamma$ $^8$Li and $^8$B decays, their 
even higher energies would allow to increase the pulse size and hence relax 
the suppression factor even further. In fact, since their end-point energies are about a factor 3.5 larger, 
one can naively expect that the number of stored ions $N^{(^8Li)}_\beta$ and $N^{(^8B)}_\beta$
can be increased by another order of magnitude compared to \neon and \he source 
beams with the same $\gamma$. For the very high $\gamma$ 
\br and \li beams, the atmospheric backgrounds are not an 
important constraint on $N_\beta$ at all, so that the only limitation on the ring design is the livetime {\em l},
\ie \, the ring geometry. 

The original $\beta$-Beam proposals assumed useful 
fluxes of $1.1\times 10^{18}$ and $2.9\times 10^{18}$ decays per year
for \neon and \he respectively
when accelerated to $\gamma=100$. 
These fluxes have been considered as ``standard'' 
and are commonly assumed in the literature.
Similar ``standard'' numbers regarding \br and 
\li are lacking. However, preliminary studies 
on the production rates of \br and \li show 
that they are both produced more easily 
than \neon and \he
respectively. This is particularly true for 
neutrinos produced from the decay of 
$^8$B since the production of $^{18}$Ne
is very challenging. Lacking any definite 
prediction for the achievable fluxes at the higher $\gamma$ considered here, 
we will assume that $10^{19}$ ions per year can be stored into the ring \cite{Donini:2008zz}, for all ion species\footnote{
For $^6$He and $^{18}$Ne boosted at $\gamma = 100$ with the original ring design proposed in Ref.~\cite{zucc}, 
with a livetime {\em l} = 0.36, this corresponds to $3.6 \times 10^{18}$ useful ion decays per year. }.

%%%%%%%%%%%%%%%%%%%%%%%%%%%%%%%%%%%%%%%%%%%%%%
\subsection{The choice of the two baselines}
\label{sec:prob}
%%%%%%%%%%%%%%%%%%%%%%%%%%%

The approximated expanded form of the expression for the golden channel probability 
keeping only up to the second order terms in the small parameters 
$\theta_{13}$ and $\ms$ \cite{golden}, can be written as in Ref.~\cite{Freund:2001pn},
\be
 P_{e\mu} &\simeq& 
 \sin^2\theta_{23} \sin^22\theta_{13}
\frac{\sin^2[(1-\hat{A})\Delta]}{(1-\hat{A})^2}\nonumber \\
&\pm& \alpha \sin2\theta_{13} \sin2\theta_{12} \sin2\theta_{23} 
\sin\deltacp \sin(\Delta) \frac{\sin(\hat{A}\Delta)}{\hat{A}}
\frac{\sin[(1-\hat{A})\Delta]}{(1-\hat{A})} \nonumber \\
&+& \alpha \sin2\theta_{13} \sin2\theta_{12} \sin2\theta_{23} 
\cos\deltacp \cos(\Delta) \frac{\sin(\hat{A}\Delta)}{\hat{A}}
\frac{\sin[(1-\hat{A})\Delta]}{(1-\hat{A})} \nonumber \\
&+& \alpha^2 \cos^2\theta_{23} \sin^22\theta_{12} 
\frac{\sin^2(\hat{A}\Delta)}{{\hat{A}}^2}
,
\label{eq:pemu}
\ee
where 
\be
\Delta\equiv \frac{\ma L}{4E_\nu},
%\ee 
%\be
~~
\hat{A} \equiv \frac{A}{\ma},
\label{eq:matt}
\ee
and $A=\pm 2\sqrt{2}G_FN_eE_\nu$ \cite{msw1,msw2,msw3} 
is the matter potential (plus/minus sign is for neutrino/antineutrino), 
given in terms of the electron density $N_e$ and 
(anti)neutrino energy $E_\nu$. 
It is easy to see from Eq. (\ref{eq:pemu}) that a way 
to get rid of all $\delta$ dependent terms is by 
considering a baseline where 
\be
\frac{\sin(\hat{A}\Delta)}{\hat{A}} = 0,
\ee
which is called the condition of the magic baseline. From 
%the line averaged density using 
the PREM density profile of the Earth \cite{prem}, this 
baseline comes out to be about $L\simeq 7000$ km. As discussed before, 
we will use this as one of our baseline options. The position of the magic baseline 
depends mainly on the density profile of the Earth and not on the oscillation parameters or the energy 
of the beam\footnote{A mild dependence on the oscillation parameters and energy 
creeps in for large values of $\theta_{13}$ \cite{magic2,betaino2}.
However, the effects are still mild.}.
However, the size of the oscillation probability does depend critically on the neutrino energy at the magic baseline.
Indeed, the density encountered by the (anti)neutrinos at this baseline allows for the denominators 
$1-\hat{A}$ in Eq. (\ref{eq:pemu}) to cancel when $E_\nu \sim 6$~GeV if the mass hierarchy is normal (inverted).
Even if the conditions under which Eq. (\ref{eq:pemu}) was expanded are, therefore, not satisfied in this case, the exact oscillation probability reveals a resonant enhancement when this condition is met \cite{betaino1}. 
The advantage of tuning the beam energy to the resonant one is two-fold: 
first, the increase in the oscillation probability compensates the loss of events due to the very long baseline, increasing the statistics at the 
far detector and improving its sensitivity to smaller values of $\sin^2 2 \theta_{13}$; second, the resonance only occurs for (anti)neutrinos if the
mass hierarchy is normal (inverted), therefore providing an extremely good probe of the mass ordering.

For the second 
baseline the most important criterion is the measurement of CP 
violation. For that we want the second term to dominate in the 
probability. Moreover, matter effects can fake true CP violation stemming from the phase $\delta$
and, therefore, short baselines and low energies are better for those studies. 
In this small matter effect regime, when $\hat{A} \to 0$ in Eq. (\ref{eq:pemu}), maximizing the CP violating terms 
amounts to require that $\sin{\Delta}=1$. For $\ma = 2.4\times 10^{-3}$ eV$^2$ this translates into $L/E=515$ km/GeV.
The mean neutrino energy of neutrinos from $^6$He and $^{18}$Ne decays at $\gamma = 350$ is $E_0 \gamma \sim 1.2$~GeV which translates
to an on-peak baseline of $L=618$~km matching perfectly the $650$~km baseline between CERN and the Canfranc laboratory. 

In the following, we will thus consider detectors located at $L  = 650$ km and $L  = 7000$ km down the source.

%%%%%%%%%%%%%%%%%%%%%%%%%%%%%%%%%%%%%%%%%%%%%%
\subsection{The Storage Ring}
%%%%%%%%%%%%%%%%%%%%%%%%%%%%%

Two geometries for the $\beta$-Beam storage ring have been considered in the
literature so far: the racetrack geometry, 
first proposed for this facility in Ref.~\cite{zucc}, and the triangle geometry,
\cite{twobaseline1}. 
 Both geometries have been considered
also in the framework of the Neutrino Factory studies, see
Ref.~\cite{Zisman:2008zz}. 

The main advantage of the triangle geometry with respect to the racetrack one
is the possibility of using, simultaneously, 
two of the three straight sections to aim to two different far detectors. 
For this reason,  a larger number of useful ion decays is achieved in triangle-shaped rings
than racetrack-shaped ones.
 Imagine now that one of the long straight sections of a triangle ring aims at
 a detector
located at $L = 7000$ km and that a second one aims at a detector located at
$L = 650$ km. 
If we inject $^6$He and $^{18}$Ne ions in the storage ring and let them decay,
neutrinos produced in the 
straight section aiming at the ``near'' detector give a very good sensitivity to
$\theta_{13}$ and to the CP violating phase
$\delta$. On the other hand, neutrinos produced in the straight section aiming
at the ``far'' detector will contribute 
scarcely to the measurement of the sign of the atmospheric mass difference,
since their energy is too small to 
have a resonant behavior in matter and compensate the very long baseline (see
Sect. \ref{sec:prob}). 
A similar situation can be observed when $^8$Li and $^8$B ions are injected
in the ring: those ions that decay aiming at the ``far'' detector produce a
neutrino flux that provides a very good sensitivity 
to the mass hierarchy, whereas those that decay in the straight section that
points to the ``near'' detector contribute 
very little to the measurement of $\theta_{13}$ and $\delta$, since the
neutrino flux is strongly off the oscillation peak, 
their energy being too high for the oscillations to develop at the 650 km baseline. 
For this reason, it is easy to understand that no particular advantage arises
in using a triangle geometry storage ring
in the set-up that we are considering. We will thus consider here two racetrack
geometry storage rings, 
each of them with one of the straight sections aiming at one of the two
detectors. Notice that this set-up is similar to the one considered in
the Neutrino Factory IDS baseline proposal \cite{Zisman:2008zz}. 

Let us now recall the main characteristics of the storage ring proposed for a
$\beta$-Beam  in the original design by Piero Zucchelli back in 2002,
\cite{zucc}. The ring was conceived to store $^6$He and $^{18}$Ne ions boosted
at $\gamma = 100$ (the maximum boost achievable using the PS and the SPS at
CERN being $\gamma = 150$ 
for $^6$He and $\gamma = 250$ for $^{18}$Ne, respectively). 
The proposed ring has a racetrack  geometry with two long straight sections of
$L_s = 2500$ m each and two arcs with curvature radius of $R =300$ m if 
a 5 T magnetic field is used to bend the ions in the curved section of the ring. 
The total length\footnote{
The length of the ring was chosen so as to match 
exactly the length of the SPS, under the assumption that this 
size was a realistic one. Notice that the ring design has not been optimized
since its first proposal.}
 of the ring  is $L_r= 2 L_s + 2 \pi R = 6885$ m, and the livetime  is {\em  l} $= L_s/L_r = 0.36$. 
The ring, with a longitudinal section of 3100 m, is tilted at
a very small angle ($\vartheta= 0.6^\circ$) with respect to the ground, so as
to aim at a detector located in the Fr\'ejus tunnel, at a distance of 130 km
from the source. 
The maximum depth $d$ of the far end of the ring with such a small tilt angle
is just $d = 32$ m. 

The original design of the ring must be modified when the boost factor
$\gamma$ is increased.  If the magnets used are LHC
dipolar magnets with a maximum magnetic field of 8.3 T, 
the curvature radius $R$ needed to bend $^6$He ions boosted at $\gamma = 350$ is $R \sim 633$ m. If the straight sections are kept
untouched, the total length of the decay 
ring becomes $L_r = 8974$ m \cite{Donini:2008zz}, whereas
the livetime decreases  to {\em l} = 0.28. Since the neutrino flux is
aimed at a detector located at 650 km from the source,  
the tilt angle in this case is $\vartheta = 3^\circ$. With a longitudinal
section of the ring of 3764 m, this means 
that the maximum depth of the far end of the ring is $d = 197$ m. 
Notice that in the same ring we can store \li ions boosted up to $\gamma = 390$ and \neon and \br ions with
$\gamma = 583$ and 656, respectively. 

It is useful at this point to compare the decay ring design proposed for a
$\beta$-Beam facility,  
depicted above, with the ring design considered in the framework of Neutrino
Factory studies, \cite{Zisman:2008zz}.  
The racetrack storage ring design for the Neutrino Factory consists of two
straight sections  of  $L_s = 600$ m each,  
with two arcs with curvature radius $R = 60$ m. The total length of the ring
is $L_r \sim 1580$ m, with a livetime  
{\em l} = 0.37. The curved sections of the ring are equipped with
superconducting dipole and quadrupole magnets. 
In the {\em International Scoping Study of a future Neutrino Factory and
  Super-Beam facility} \cite{Bandyopadhyay:2007kx}, two distances have emerged
as optimal locations for far detectors: $L \sim 3500$ km and the magic baseline, $L \sim 7500$
km. The tilt angle to aim at these two baselines are 
$\vartheta = 16^\circ$ and $\vartheta = 36^\circ$, respectively. Since the
longitudinal section of the storage ring  
is 720 m, the maximum depth at the far end of the ring is $d = 198$ m for the
$L  = 3500$ km baseline  
and $d = 423$ m for the $L = 7500$ km one. Notice that the Neutrino Factory
racetrack geometry ring is much  
more compact than the analogous device proposed for the $\beta$-Beam. The
different size is motivated by two important differences between the
$\beta$-Beam and the Neutrino Factory: first,  shorter arcs are needed to bend
muons with  
respect to ions, for similar magnetic fields; second,  the occupancy of a
$\beta$-Beam ring must be very small to reduce the atmospheric background
as stressed at the beginning of this section ({\it i.e.}, either we inject very few ions into the ring, 
or  the size of the ring must be very large) .
The atmospheric background, however, is not a significant
problem at the Neutrino Factory\footnote{ 
Notice that the so-called ``{\it low-energy Neutrino Factory}'', proposed in
Ref.~\cite{Geer:2007kn,Bross:2007ts},  
could be affected by the same problem as the $\beta$-Beam.  In this case, the
storage ring design for this  
facility should be modified accordingly.}, the neutrino energy being of the
order of several GeVs (in this range  
of energy the atmospheric background is at least two orders of magnitude
smaller than in the case of $O(100)$ MeV neutrinos).  

From the comparison with the Neutrino Factory ISS/IDS study,  it emerges that the
original design by Piero Zucchelli 
for a racetrack ring aiming at $L = 650$ km (modified to take into account the
higher ions boost factor) is not 
unrealistic: albeit longer than the ring conceived for the Neutrino Factory,
the decay tunnel for this ring reaches 
the same depth $d$ as the Neutrino Factory ring aiming at $L = 3500$ km. 
However, if a ring of the same type is used to aim at a detector located at $L= 7000$ km from the source, 
the tilt angle to be considered is $\vartheta = 34.5^\circ$. In this case, the
maximum depth of the far end of the  
ring is $d = 2132$ m, something well beyond any realistic possibility. 
As it was stressed in the beginning of this section, however, two storage
rings will be used to aim to the detectors located at $L = 650$ km and $L = 7000$ km. 
Therefore, it is possible to design two rings of different 
characteristics, each of them optimized for a different detector. 
In particular, the ring aiming at the magic baseline could be more compact than the other one. 
One possibility is to use the slightly more favorable $Z/A$ ratio of \li with respect to \he to build 
a ring with curvature radius $R \sim 562$ m, $L_r = 8531$ m and maximal depth $d = 2053$ m.
It is clear that the gain achievable with this option is not significant, although the livetime increase to {\em l} = 0.29. 
A second, more interesting, possibility is to reduce the straight sections of the ring to reduce its longitudinal size, and correspondingly $d$,
at the price of a reduced livetime. A relevant question is, then, how much can we reduce the livetime  
of the ring so as to increase its technical feasibility, but with only a small loss in the sensitivity to the mass hierarchy? 
Even more important, which loss of sensitivity to the mass hierarchy is acceptable without a significant loss  
of sensitivity to the CP violating phase $\delta$? 

An answer to these questions is offered by Table \ref{tab:rates}. We can see that increasing the boosting factor of \li and \br ions to the maximum $\gamma$ for which these ions can still be stored into a ring with $R \sim 633$ m, 
a significant increase of the number of events in the far detector can be achieved. Such increase depends on the hierarchy
and on the fulfillment of the resonant condition of the oscillation probability in matter: for example, a 10\% increase
of the boost of \li ions from $\gamma = 350$ to $\gamma = 390$ implies a 50\% (25\%) increase in the number of events observed at the detector for inverted (normal) hierarchy. Similar results are obtained for \br ions. 

\begin{table}[hbtp]
\renewcommand{\arraystretch}{1.7}

\begin{center}

\begin{tabular}{|c|c|c|c|c|c|c|}
\hline 
  & $\gamma^{^8Li}$ & 350 & 360 & 370  &  380 &  390  \\
\hline
\hline
% -----------------------------------------------------------------------------
%
\multirow{2}{*}{+ } & $ N_{ev} (\gamma)$                  & 1.84 & 1.94  & 2.05 & 2.18  & 2.33 \cr
\cline{2-7}$  $ & \,$N_{ev} (\gamma)/N_{ev} (350)$\,  &          & 1.05  & 1.11 & 1.18  & 1.27 \\ 
\hline\hline
 \multirow{2}{*}{- } &  $N_{ev} (\gamma)$ & 55.80 & 62.46 & 69.40 & 76.54 & 83.86 \cr
  \cline{2-7} & \,$N_{ev} (\gamma)/N_{ev} (350)$\, & & 1.12  & 1.24 & $1.37 $ & 1.50 \\ 
\hline
\hline
  & $\gamma^{^8B}$ & 583 & 600 & 617 &  633 &  650  \\
\hline
\hline
% -----------------------------------------------------------------------------
%
\multirow{2}{*}{+ } & $ N_{ev} (\gamma)$                  & \,477.16\, & \,499.72\,  & \,521.64\, & \,541.68\, & \,562.34\, \cr
\cline{2-7}$  $ & \,$N_{ev} (\gamma)/N_{ev} (583)$\,  &     & 1.05   & 1.09 & 1.14  & 1.18 \\ 
\hline\hline
 \multirow{2}{*}{- } &  $N_{ev} (\gamma)$                  & 15.20 &  16.58  & 17.99 & 19.34 & 20.79 \cr
  \cline{2-7} & \,$N_{ev} (\gamma)/N_{ev} (583)$\,      &  &  1.09   & 1.18 &  1.27  & 1.37 \\ 
\hline
\hline
\end{tabular}
\end{center}
\caption{\label{tab:rates} Number of muons observed at a 50 kton magnetized iron detector \cite{ino} 
with perfect efficiency located at 7000 km from the source after 5 years of data taking
 as a function of the boost factor of \li (upper table) and \br ions (lower table), for $\theta_{13} = 5^\circ$ and $\delta = 90^\circ$. A livetime $l=0.3$ was also assumed for the storage ring. The ratio of the number of events 
obtained with a given $\gamma$ with respect to those obtained storing \li (\br) ions boosted at 
 $\gamma = 350$ (583) is also shown.}
\end{table}

The increase in the statistics can be used for two different purposes: the
first possibility, of course, is to use  
it to achieve a higher sensitivity to the mass hierarchy. However, 
the sensitivity increase is not dramatic (as it should be expected, since for
Gaussian statistics the sensitivity 
scales with the square root of the statistics). The second possibility, that
could open the path to a feasible  
$\beta$-Beam facility with long baseline, is to use the higher statistics to
reduce significantly the size of the storage ring:  
the physics reach of a set-up with a racetrack ring with $L_r = 8531$ m and
{\em l} = 0.29 (described above) 
with $^8$Li ions boosted at $\gamma = 350$ is identical to the reach of a
racetrack ring with a much shorter straight section, 
$L_s = 998$ m, if the $^8$Li ions are boosted at $\gamma = 390$. This ring
has a total length $L_r = 5970$ m, a  
longitudinal section of 2263 m and a livetime {\em l} = $0.6 \times 0.28 \sim 0.17$. 
The maximum depth  of the far end of this ring is $d = 1282$ m. Such a depth is
still much larger than what is needed for the Neutrino Factory rings 
(we remind that $d = 423$ m is the maximum depth of the far end of the ring aiming  
at $L = 7500$ km), but is almost 1 km shorter than for the standard design of the ring.  
Note that for the higher energy $^8$Li/$^8$B beams, the problem of atmospheric neutrino background is 
almost non-existent, as discussed before. Therefore, the reduction of the total ring size 
does not pose any serious threat to the experiment. 

\vglue 0.3cm
\noindent
We therefore propose a \bb set-up with two storage rings of different design: 
\begin{itemize}
\item One ring for the $^6$He/$^{18}$Ne 
ions with $l=0.28$, sending the beam to $L=650$ km (to Canfranc, Spain).  Both \he and \neon ions are
boosted at $\gamma = 350$ (no significant gain is achieved by boosting \neon ions to higher $\gamma$'s);
\item A second ring  for the $^8$Li/$^8$B ions with 
$l=0.17$, sending the beam to $L=7000$ km (to INO, India). In this case, \li ions are boosted at $\gamma = 390$
and \br ions at $\gamma = 656$ ({\em i.e.}, the maximum $\gamma$ that permits to store the ions in a ring with 8.3 T magnets).
\end{itemize} 
Both rings have curvature radius $R = 633$ m, with straight sections of length $L_s = 2500$ m  and 998 m, respectively.
The maximal depth at the far end of each ring is $d = 197$ m for the 650 km baseline and $d = 1282$ m for the 
magic baseline.

A more compact ring (with a higher {\em l})
could be obtained by increasing the magnetic field in the curved section, taking  
advantage of the R\&D programme  for LHC upgrades aimed to the development of
high field magnets (with $B  \in [11-15]$ T). 
If one assumes that magnetic field strengths of 15 T could be used for the storage ring \cite{Donini:2008zz}, 
then \he ions boosted at $\gamma = 350$ could be stored in a ring with curvature radius $R = 350$ m. 
If the straight sections of the ring 
are kept fixed to $L_s = 2500$ m, the total length of the ring is $L_r = 7200$ m with a livetime $l = 0.35$. 
The longitudinal section of this ring would be 3200 m, with a maximal depth $d$ at the far end of the ring when tilted at 
$\varphi = 34.5^\circ$ of 1812 m. If we now fill a ring equipped with the same magnets with \li and \br
ions boosted at $\gamma=390/656$, we can still achieve a good sensitivity to the mass hierarchy 
reducing the livetime to $l = 0.17$ (as discussed above), corresponding to straight sections of length $L_s = 556$ m. 
Such a ring has a total length $L_r = 3311$ m, a longitudinal section of 1256 m and a maximal depth at the 
far end of the ring aiming at the magic baseline detector $d = 711$ m.
This depth is not much larger than the depth required for the Neutrino Factory magic baseline ring, 
and hence it could represent an extremely interesting option to be investigated further.

%%%%%%%%%%%%%%%%%%%%%%%%%%%%%%%%%%%%%%%%%%%%%%
\subsection{The Detectors}
%%%%%%%%%%%%%%%%%%%%%%%%%
Unlike the Neutrino Factory, or the Super-Beams, the $\beta$-Beam is a truly mono-flavor neutrino beam. 
Therefore, while for the detector of the Neutrino Factory beam charge 
identification capability is mandatory in order to 
tag the initial neutrino flavor, 
this needs not be the case for $\beta$-Beams. The only 
criterion is that the detector should have a good particle identification 
sensitivity, and in particular should be able to distinguish a muon 
from an electron. 
Most known detector technologies have been 
considered in the literature for this class of 
experiment. 
%In particular, water \chr detectors have been 
%considered in \cite{wc}, 
%TASD in \cite{scin,twobaseline2}, and 
%magnetized iron in \cite{mag,paper1,betaino1,betaino2}. 
Each of these 
detectors offer the best performance for only a certain energy 
range of the neutrinos. A detailed report card on the 
detector performance in terms of energy threshold, energy 
resolution, backgrounds, statistics and costs is required 
for deciding the best detector option. The detector choice is 
also directly dictated by the energy of the $\beta$-Beam. 

%%%%%%%%%%%%%%%%%%%%%%%%%%%%%%%%%%%%%%
\begin{table}
\begin{center}
\begin{tabular}{||c||c||c||c||} \hline\hline
{\rule[0mm]{0mm}{6mm}\multirow{2}{*}{Detector Characteristics}}
&{\rule[-3mm]{0mm}{6mm}{MIND \cite{ino,Abe:2007bi}}}
& {\rule[-3mm]{0mm}{6mm}{TASD \cite{nova}} }
& {\rule[-3mm]{0mm}{6mm}{WC \cite{bc}}}
\cr
& (Only $\mu^{\pm}$) & (Both $\mu^{\pm}$ \& $e^{\pm}$) &\cr
\hline\hline
Fiducial Mass & 50 kton & 50 kton & 500 kton \cr
\hline\hline
$E_{min}$ & 1 GeV & 0.5 GeV & 0.5 GeV \cr
\hline
$E_{max}$ & 18 GeV & 2.5 GeV & 2.5 GeV \cr
\hline
Bin Size & $\in\left[ 0.6,\, 2.3 \right]$ GeV & 0.2 GeV & 0.25, 0.5 GeV \cr
\hline\hline
%Energy Resolution function ($\sigma$)& 0.15E & 0.03$\sqrt E$ ($P_{e\mu}$) \& 0.06$\sqrt E$ ($P_{ee}$)\\[2mm]
Background Rejection & 0.0001 & 0.001 & $\in\left[ 0.0001,\, 0.001 \right]$ \cr
\hline
Signal error (syst.) & 2.5\% & 2.5\% & 2.5\% \cr
\hline
Background error (syst.) & 5\% & 5\% & 5\% \cr
\hline\hline
Detection Efficiency ($\epsilon$) & $\in\left[ 5,\,70 \right]\%$ & 80\% ($\mu^{\pm}$) 
\& 20\% ($e^{\pm}$)& $\in\left[ 20,\, 50 \right]\%$ \cr
\hline
\multirow{2}{*}{Energy Resolution ($\sigma$) }& \multirow{2}{*}{0.15 E(GeV)} & 0.03$\sqrt{\rm E
(GeV)}$ for $\mu^{\pm}$ & \multirow{2}{*}{$\lesssim$0.15E(GeV)} \cr
                      &   &  0.06$\sqrt{\rm E (GeV)}$ for $e^{\pm}$ &\cr
\hline
Charge Id Efficiency ($f_{ID}$)& Yes & No & No \cr
\hline\hline
\end{tabular}
\caption{\label{tab:detector}
Comparison of the typical 
detector characteristics expected for the three most 
popular $\beta$-Beam detectors. 
}
\end{center}
\end{table}
%%%%%%%%%%%%%%%%%%%%%%%%%%%%%%%%%%%%%%%%%%%%%%%%%%%%%%

For the \neon and \he $\beta$-Beam, it was argued in Ref.~\cite{betaoptim}
that the water \chr detector would be best for $\gamma \ltap 300$, 
while for larger boost factors one should use the TASD detector. 
In fact, most studies have used 
%with Beta-Beams have considered the low to intermediate energy 
%beam and have hence used 
megaton scale water \chr detectors as detector option for a $L\leq 1000$ km
\cite{oldpapers,donini130,bc,bc2,doninialter,fnal,boulby}. 
In Refs.~\cite{paper1,doninibeta} the idea of observing 
high $\gamma$ $\beta$-Beam
neutrinos with magnetized iron detectors was introduced for the 
first time. This prospect was further perused in 
Refs.~\cite{betaino1,betaino2,rparity} and later in 
Refs.\cite{twobaseline1,bboptim,twobaseline2}. 
We show in Table \ref{tab:detector} the comparative catalogue of 
detector characteristics. The first relevant difference between the different technologies
is the energy threshold: both the TASD \cite{nova} and water \chr detectors \cite{bc}
have a very low energy threshold and are, hence, ideal for neutrino beams of relatively low energy (up to a few
GeV). Magnetized iron detectors of the MIND type \cite{Abe:2007bi}  (see also \cite{ino}), on the other hand, 
are a good option only for higher energy beams. 
The energy resolution of TASD is impressive up to a few GeV, whereas that
of water \chr detector is good, but only for the energy regime which has a predominance of quasi-elastic 
events ($E \ltap 1$ GeV). Eventually, iron detectors energy resolution is limited by the present segmentation design.
The background rejection fraction, on the other hand, is seen to be best for the magnetized iron detector. 
It is in fact expected to be better for magnetized iron by at least an order of magnitude compared to water \chr and TASD. 
Scaling of the detector mass is difficult for TASD and magnetized iron detectors beyond 50 kton, whereas
megaton scale water \chr detectors are  currently under study \cite{t2k,Memphys}.
Notice, however, that a 50 kton magnetized iron detector represents, at present, the cheapest option between 
the three detectors technologies and design considered in Table \ref{tab:detector}.

Based on the comparative performance of the detectors and our 
physics goals we make the following choices: (1) Since the shorter baseline is the optimal one to 
perform CP violation studies, and since CP measurements are better at lower energies with \neon and \he 
as source ions than at higher energy with \li and \br \cite{twobaseline1,twobaseline2}, it is preferable to have 
a detector with lower threshold and good energy resolution. Therefore, the choice would be between TASD and water \chr detectors. Since the latter can be made larger than TASD, we opt for a 
water \chr detector with 500 kton fiducial mass at  the shorter baseline (as in Refs.~\cite{bc,bc2}). 
This detector could be housed at Canfranc, for example, at a distance of 650 km from the $\beta$-Beam at CERN; 
(2) Mass hierarchy measurement is the main motivation for the experiment at the magic baseline, for which higher energy neutrinos from highly boosted \br and \li ions will be used. We prefer thus to use a magnetized iron detector
at this baseline. This far detector could be the ICAL@INO detector in India \cite{ino} which is at a distance of 
7152 km, tantalizing close to the magic baseline, and which will soon go under construction. 
We will assume 50 kton of detector mass for this case, though it is possible that INO will be 
upgraded to 100 kton. Notice that the numerical analysis has been performed for a baseline $L = 7000$ km.

In order to simulate the response of the water \chr and magnetized iron
detectors when exposed to the $\beta$-Beam fluxes, we follow
the analyses performed in Refs. \cite{bc2} and \cite{Abe:2007bi}. The
efficiencies and beam-induced backgrounds expected in a 
water \chr detector for the $\gamma = 350$ $\beta$-Beam fluxes from \neon and
\he decays are given in \cite{bc2} as migration matrices that we use to
simulate our ``near" detector. Unfortunately, a similarly detailed analysis of
the performance of the iron detector exposed to the $\beta$-Beam fluxes is
lacking. We therefore follow the efficiencies and backgrounds derived in
\cite{Abe:2007bi} for the Neutrino Factory fluxes instead (see, also, Ref.~\cite{mind}). 
Notice that this is a very conservative assumption since charge ID is not mandatory in a
$\beta$-Beam, unlike for the Neutrino Factory, given the purity of the
beam. Moreover, the Neutrino Factory spectrum is much wider than the $\beta$-Beam one and reaches
much higher energies. Higher energy events, in turn,  can induce neutral current
interactions that feed down background to lower energies. The largest
uncertainties in the performance of the iron detector are on the efficiencies
and backgrounds for the events of lowest energy, around $1-5$ GeV. However,
the main role of the iron detector considered in this set-up is to observe the
resonant enhancement of the oscillation probability that happens around $6-7$
GeV to measure the mass hierarchy. Therefore, the performance of the proposed
set-up does not depend critically on the efficiency and background of the
lowest energy events, unlike in the Neutrino Factory IDS baseline design where
these events are crucial to solve degeneracies and improve the sensitivity to
CP violation for large $\theta_{13}$. We will illustrate the mild dependence
of the performance of the set-up on the energy threshold of the detector in
the next section.

%%%%%%%%%%%%%%%%%%%%%%%%%%%%%%%%%%%%%%%%%%%%
\section{Comparative Sensitivity Reach}
%%%%%%%%%%%%%%%%%%%%
\label{sec:results}
%%%%%%%%%%%%%%%%%%%%%%%%%%%%%%%%%%%%%%%%%%%%%%%%%%%%%%%%

In this section we probe the sensitivity reach of 
the $\beta$-Beam set-up that we have defined in the 
previous section. 
We are interested in 
looking at the performance of a given experiment to discover 
$\theta_{13}$, CP violation, and the mass hierarchy. 
We therefore 
quantify the sensitivity reach of the experiments in terms of 
three different performance indicators. 
\begin{enumerate}
\item The $\stch$ discovery reach: This is the minimum 
true value of $\stch$ for which the experiment can rule out at 
$3\sigma$ 1 d.o.f. the value $\stch=0$ in the fit, after marginalizing over all the other parameters. 
This gives the limiting true value of $\stch$ for which the data can statistically 
distinguish a positive $\theta_{13}$-driven oscillation from 
the $\theta_{13}=0$ prediction. 
\item The CP violation 
reach: This is the range of $\delta$ as a function of 
$\stch$ which can rule out no CP violation ($\delta=0$ and 
$180^\circ$) at $3\sigma$ 1 d.o.f., after marginalizing over all the other parameters. 
\item The $sgn(\ma)$ reach in $\stch$: This is defined as the 
limiting value of $\stch$ for which the wrong hierarchy can be 
eliminated at $3\sigma$. Below this value of 
$\stch$, the predictions for the wrong hierarchy cannot 
be separated from the data corresponding to the right hierarchy, 
at a statistical significance of $3\sigma$. We will show these 
results for both normal and inverted hierarchies.
\end{enumerate}

\begin{figure}
\includegraphics[width=0.46\textwidth,angle=0]{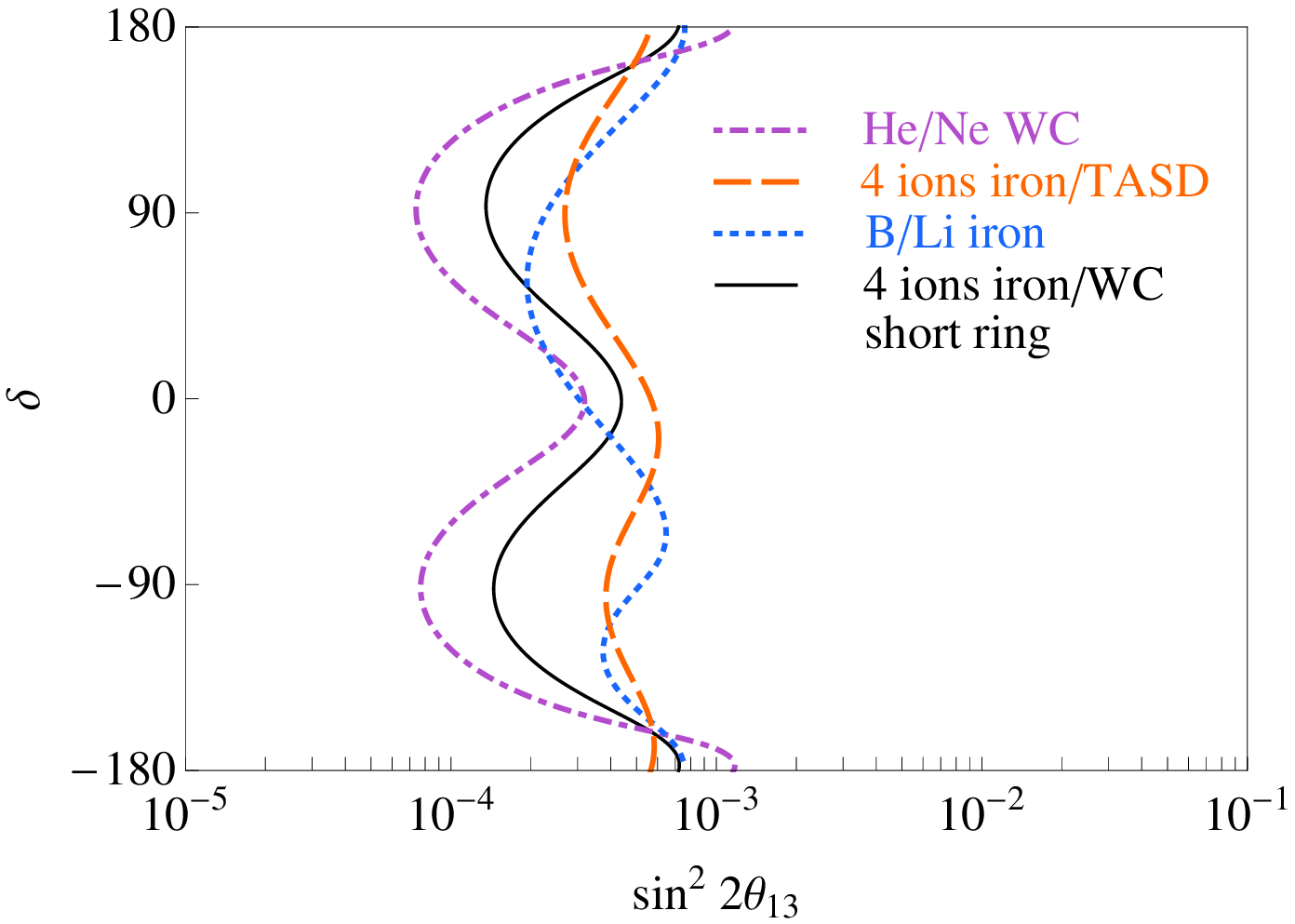}
\hglue 0.2cm
\includegraphics[width=0.46\textwidth,angle=0]{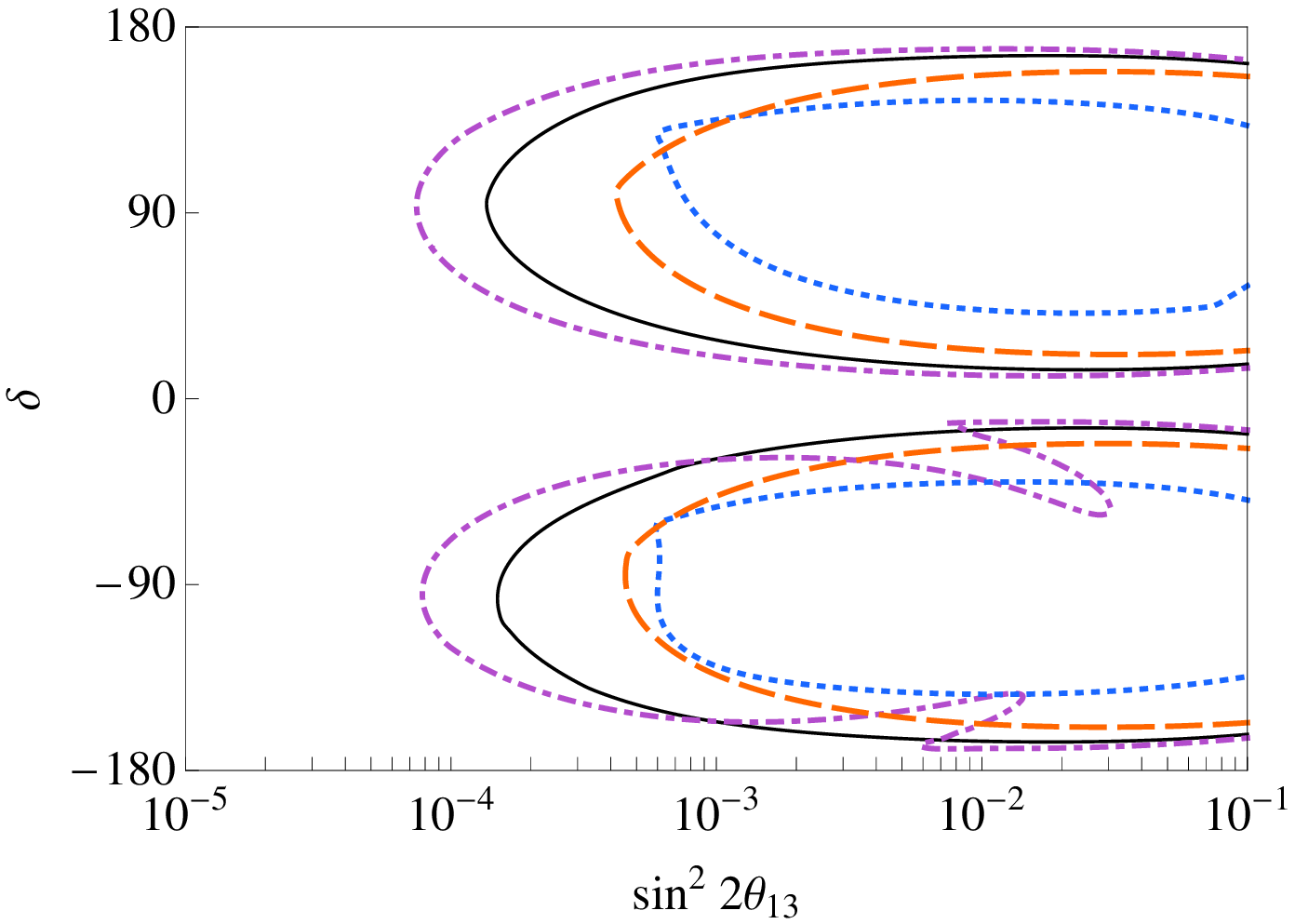}
\vglue 0.2cm
\includegraphics[width=0.46\textwidth,angle=0]{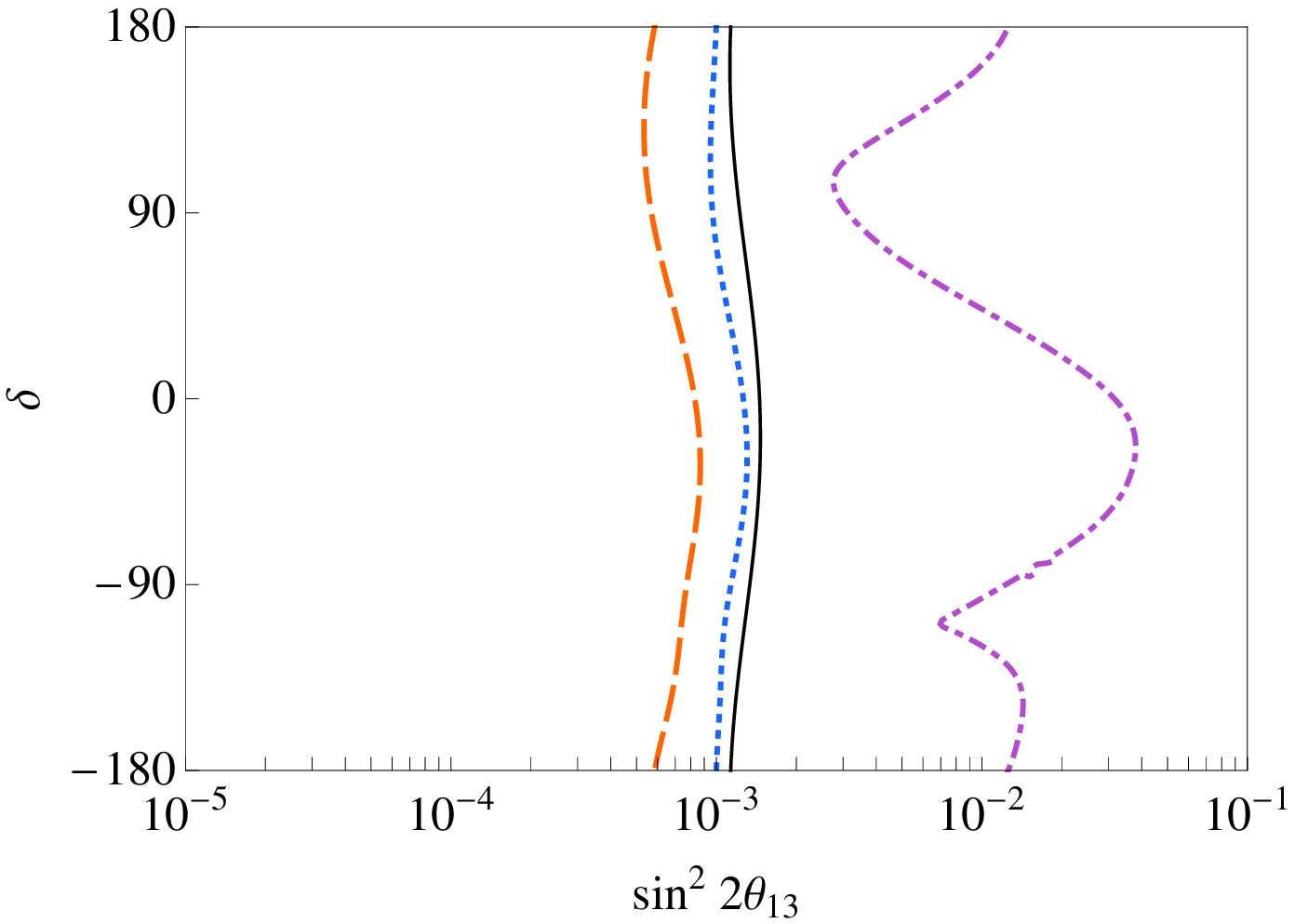}
\hglue 0.2cm
\includegraphics[width=0.46\textwidth,angle=0]{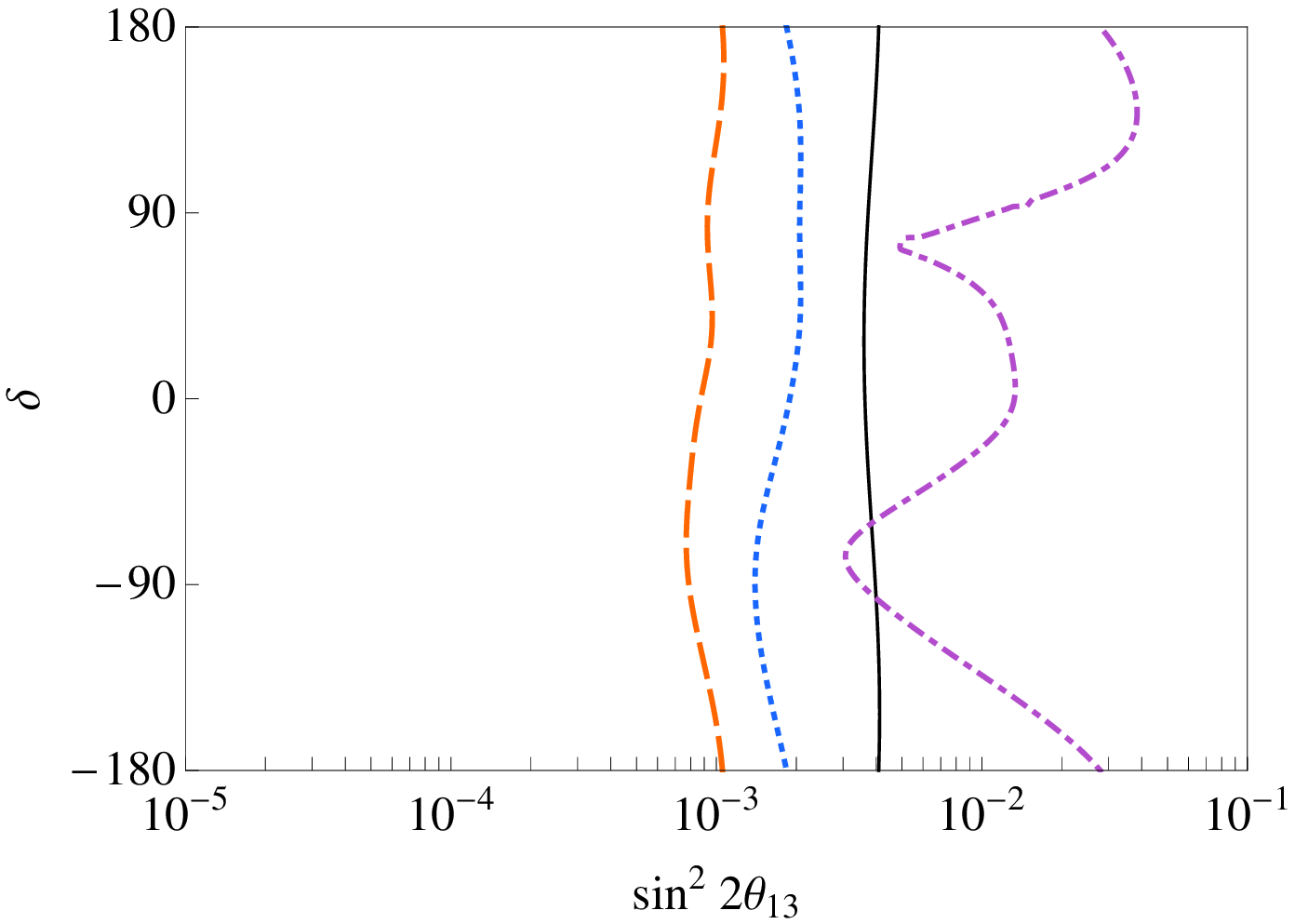}
\caption{\label{fig:sens}
Sensitivity reach of the different \bb set-ups in terms of the 
three performance indicators defined in the text. 
The upper left hand panel 
shows $\stch$ discovery reach, the upper right hand 
panels shows the CP violation reach, while the lower 
panels show the mass hierarchy discovery reach. The 
lower left hand panel is for normal hierarchy as true 
while the lower right hand panel shows the corresponding 
reach when inverted hierarchy is true. The different line types 
are for different \bb set-ups as described in the text. Note that the 
black lines are for the set-up proposed in this paper and 
has the \br and \li storage ring which has straight sections 
shorter by 60\% compared to all other set-ups (\ie, a 40\% smaller flux at the far detector).}
\end{figure}

\noindent
In Fig. \ref{fig:sens} the black lines show 
the sensitivity reach of our proposed set-up 
in terms of the three performance indicators 
defined above. 
We also compare its performance with three other high 
$\gamma$ $\beta$-Beam set-ups, the sensitivity reaches 
for which are also shown. To make a fair comparison, 
we (re)calculate the sensitivities for each of the 
benchmark set-ups assuming the same total 
number of radioactive ions injected in the storage ring(s) 
and the same total number of years of running of the experiment. 
We assume that, at a given time, only one source ion is 
accelerated and fed into a storage ring. 
Expected performance of each of these benchmark set-ups 
is shown by a particular line type, 
and they are defined as follows: 

\begin{enumerate}
\item Solid, black lines: This corresponds to the 
two-baseline $\beta$-Beam 
set-up proposed in
  this paper. Neutrino beams produced by \neon and \he decays, 
each accelerated to $\gamma =
  350$ and detected in a 500 kton water \chr detector 
located at 650 km. A second set of 
  beams from \br and \li decays with $\gamma = 656$ and $\gamma = 390$,
  respectively, are detected at 7000 km by a 50 kton magnetized iron detector.
The straight sections of storage ring of the \br and \li 
source ions are 60\% {\it shorter}  than in the original ring design, and 
the total \br and \li fluxes at the far detector is 40\% smaller. 
\item Blue, dotted lines: The two-baseline \bb set-up proposed in 
\cite{twobaseline1}. Here neutrino beams from decay of \br and \li 
with boost factor $\gamma = 350$, are detected in 
two 50 kton magnetized iron detector
located at 2000 km and 7000 km respectively.
\item Orange, dashed lines: The two-baseline \bb set-up proposed in 
\cite{twobaseline2}. Here all four ions are used. Beams 
from decays of \neon and \he accelerated to $\gamma = 575$ 
are detected in a 50 kton TASD detector at 730 km. 
Beams from decays of \br and \li accelerated to $\gamma = 656$ 
are detected in a 50 kton magnetized iron detector 
at 7000 km. 
\item Purple, dot-dashed lines: The one-baseline 
\bb set-up proposed in \cite{bc,bc2}. 
Neutrino beams produced by \neon and \he decays, 
each accelerated to $\gamma =
350$ are detected in a 500 kton water \chr detector 
located at 650 km. 
\end{enumerate}

\noindent
For all the four set-ups we assume that 
there are {\it $10^{19}$ total decays per year}, irrespective 
of the choice of the ion \cite{Donini:2008zz}. 
Of these, only ions which decay along the straight section of the 
storage ring aimed at one of the two detectors are useful. 
For the ``standard'' storage ring considered in set-ups 2, 3 and 4, the livetime is {\em l} = 0.28. 
We have, thus, used {\it $3 \times 10^{18}$ useful decays per year} 
for each ion species to reproduce the reach to the three observables for these earlier proposals.
However, for the proposal made in this paper, the storage ring for the \br and \li ions 
have straight sections which are shorter by 60\%, giving a livetime
that is 40\% smaller than for the standard storage ring. 
Accordingly, for the \br and \li generated fluxes, we have only 
{\it $0.6 \times 3 \times 10^{18}$ useful decays per year}. 
We conservatively assume that only one type ion can be accelerated 
at a time and consider a
total runtime of 10 years for all the set-ups we compare. 
We thus consider 5 years run per source ion for the experiments 
with two ions\footnote{For set-up 2 where we have two ions
but two baselines, we are therefore assuming that both detectors 
are irradiated simultaneously with neutrino beams from each ion for 
5 years each. This can be done, as suggested in Ref.~\cite{twobaseline1}, 
using a triangular geometry storage ring, with a total livetime {\em l} = 0.46, 
\ie with a flux aimed at each detector of $0.23 \times 10^{19}$ useful decays per year.}, 
and 2.5 years run per ion for those with four ions.
We have considered 2.5\% and 5\% systematic errors on the 
signal and on the beam-induced background, respectively. 
They have been included as ``pulls'' in the statistical 
$\chi^2$ analysis. The following $1 \sigma$ errors for 
the oscillation parameters were also considered: 
$\delta \theta_{12} = 1\%$, $\delta \theta_{23} = 5\%$, 
$\delta \Delta m^2_{21} = 1 \%$ and $\Delta m^2_{31} = 2\%$. 
Eventually, an error $\delta A = 5\%$ has been considered for the Earth
density given by the PREM model \cite{prem}. 
Marginalization over these parameters 
has been performed for all observables. The Globes 
3.0 \cite{globes1,globes2} software was 
used to perform the numerical analysis.

The upper left hand panel of Fig. \ref{fig:sens} 
shows the $\stch$ discovery reach. As it can be seen, the four set-ups 
perform in a very similar way. 
While for particular values of $\delta\simeq \pm90^\circ$,  
the best reach comes from set-up 4, with $^6$He/$^{18}$Ne 
ions and water \chr detector (purple dot-dashed line), $\stch \leq 7 \times 10^{-5}$, 
its $\delta$-marginalized sensitivity is seen to be the poorest. 
This happens due to the very strong $\delta$-dependence of 
the probability at $L=650$ km. On the other hand, the two 
baseline set-ups 2 (blue dotted line) and 3 (orange dashed line)
which involve the magic baseline as well, show very little $\delta$-dependence. 
The set-up proposed in this paper (black solid line), 
apparently shows some $\delta$-dependence despite having one of the 
detectors at the magic baseline because the near detector  
in this case is 10 times larger than the near detectors 
for set-ups 2 and 3. Therefore, while the $\delta$-marginalized 
$\stch$ discovery reach of our proposed set-up is similar to  
that for both the earlier two baseline set-ups, we see more 
$\delta$-dependence here due to the 10 times larger detector 
at the shorter baseline. 
Note that while the flux 
is comparatively lower at the magic baseline, 
the probability is higher. The latter therefore compensates 
the effect of the former and we expect the same statistics 
per kton of the detector at both baselines. 
However, the detector size for 
water \chr has been taken as 10 times larger compared to magnetized 
iron or TASD. Therefore, the statistics at the 
water \chr detector at $L=650$ km is  
10 times larger compared to the statistics at the 
magnetized iron detector at $L=7000$ km. For this reason, 
the results of set-up 1 follows closely those of set-up 4: 
the ultimate $\stch$ reach for our setup $\stch \leq 2 \times 10^{-4}$, is 
also obtained for $\delta \simeq \pm 90^\circ$.

The upper right hand panel shows the CP violation discovery potential. This is best at the shorter
baselines. Thus, the facilities with larger number of events at short baseline 
outperform the others in their CP violation reach. This means that set-up 4,
from \cite{bc,bc2} has sensitivity to CP violation for the smallest values of 
$\stch$, since the short baseline water \chr detector is exposed to the beam for ten years
(\ie, all the considered runtime). Unsolved sign degeneracies due to the lack of events at longer baselines, however, 
spoil the sensitivity for negative values of $\delta$ around $\sin^2
2 \theta_{13}\sim 10^{-2}$ (the so-called ``$\pi$-transit'' \cite{Huber:2002mx}).
This problem is solved when a magic baseline detector is added to the on-peak one. 
For this reason, no loss in the discovery potential is found for set-ups 1, 2 and 3 for 
particular values of $\theta_{13}$. Notice that the set-up that we propose in 
this paper has the next-to best performance (the near detector is exposed to the
beam for five years instead of ten) and no $\pi$-transit problem. 
Finally ,the worst performance for CP violation is that of the set-ups 2 and 3, in which the near detector
has a fiducial mass of 50 kton, only.

The lower panels show the sensitivity to the mass hierarchy. 
This is best at the far detectors and thus, the facilities with larger number of events 
at the magic baseline perform best. That explains the much smaller sensitivity 
of set-up 4 from \cite{bc,bc2} with no events at the longer baseline. 
The best sensitivities are in this case achieved 
for set-up 3 from \cite{twobaseline2} due to the higher statistics granted by
the larger gamma factor assumed of $\gamma = 656$ for both \br and \li. This
plots shows the advantage of accelerating the ions to higher energies. Since
for the set-up we propose here we restrict to the maximum $\gamma$ attainable
at the SPS+, which for \li is $\gamma = 390$, the difference between the two
set-ups is larger for the inverted hierarchy (lower right hand panel), where
the sensitivity stems mainly from the antineutrinos from \li decays.
The ultimate sensitivity to the mass hierarchy for our set-up is $\stch \leq 1 \times 10^{-3} (4 \times 10^{-3})$
for normal (inverted) hierarchy, independently from $\delta$. This must be compared with 
$\stch \leq 6 \times 10^{-4} (1 \times 10^{-3})$ for normal (inverted) hierarchy, achievable with set-up 3 \cite{twobaseline2}.

\subsection{Detector and decay ring specification dependence}
\begin{figure}
\includegraphics[width=0.46\textwidth,angle=0]{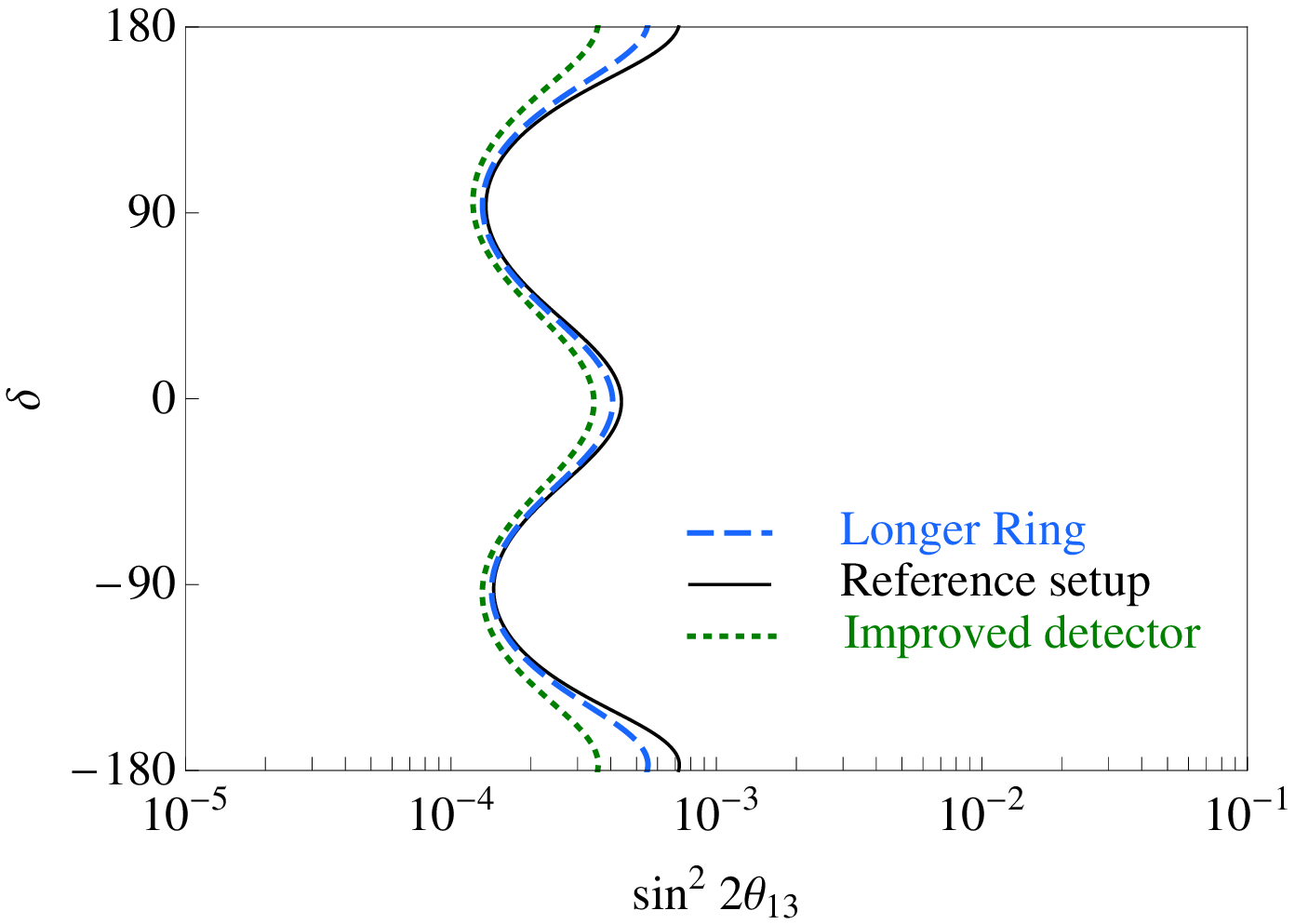}
\hglue 0.2cm
\includegraphics[width=0.46\textwidth,angle=0]{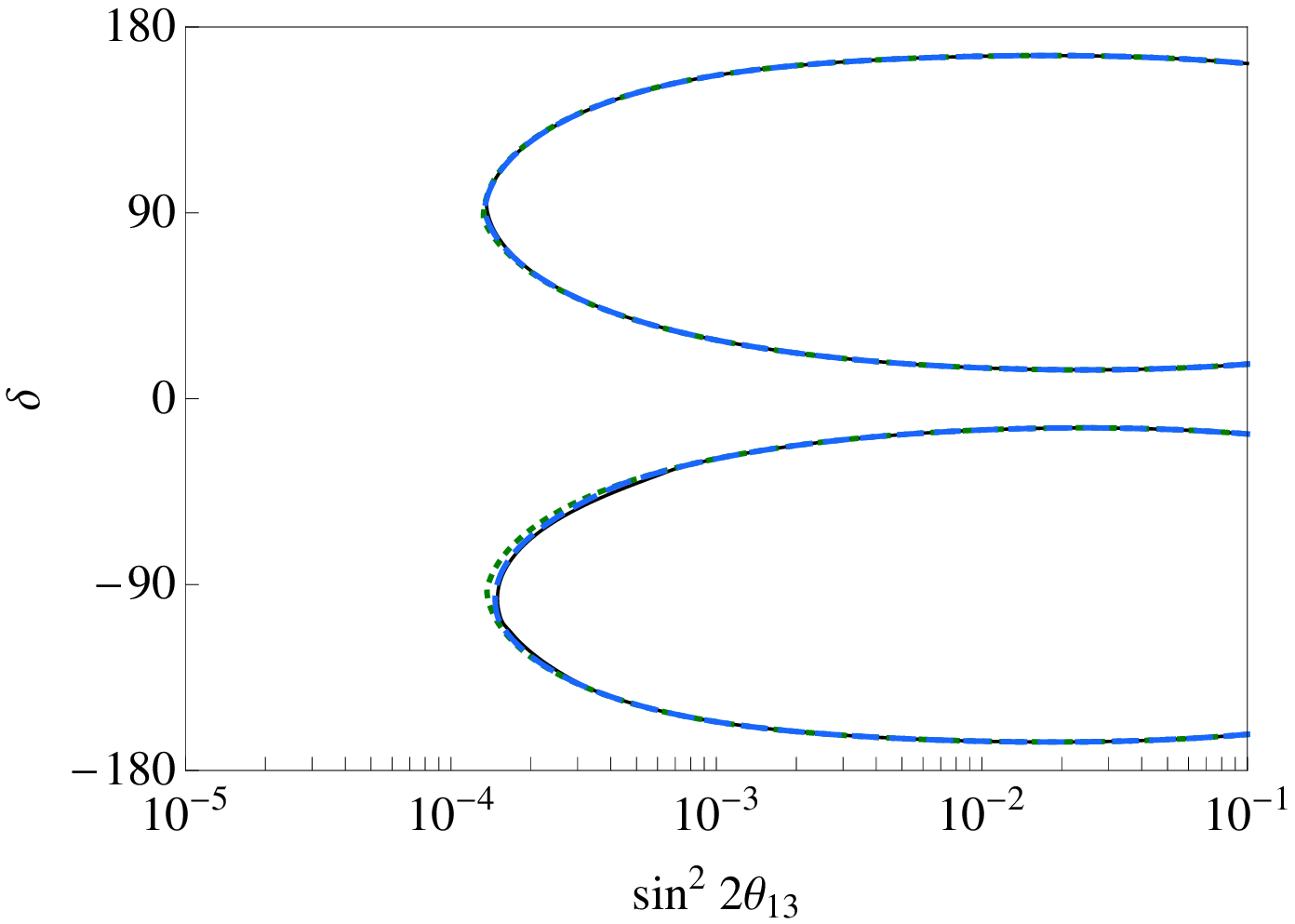}
\vglue 0.2cm
\includegraphics[width=0.46\textwidth,angle=0]{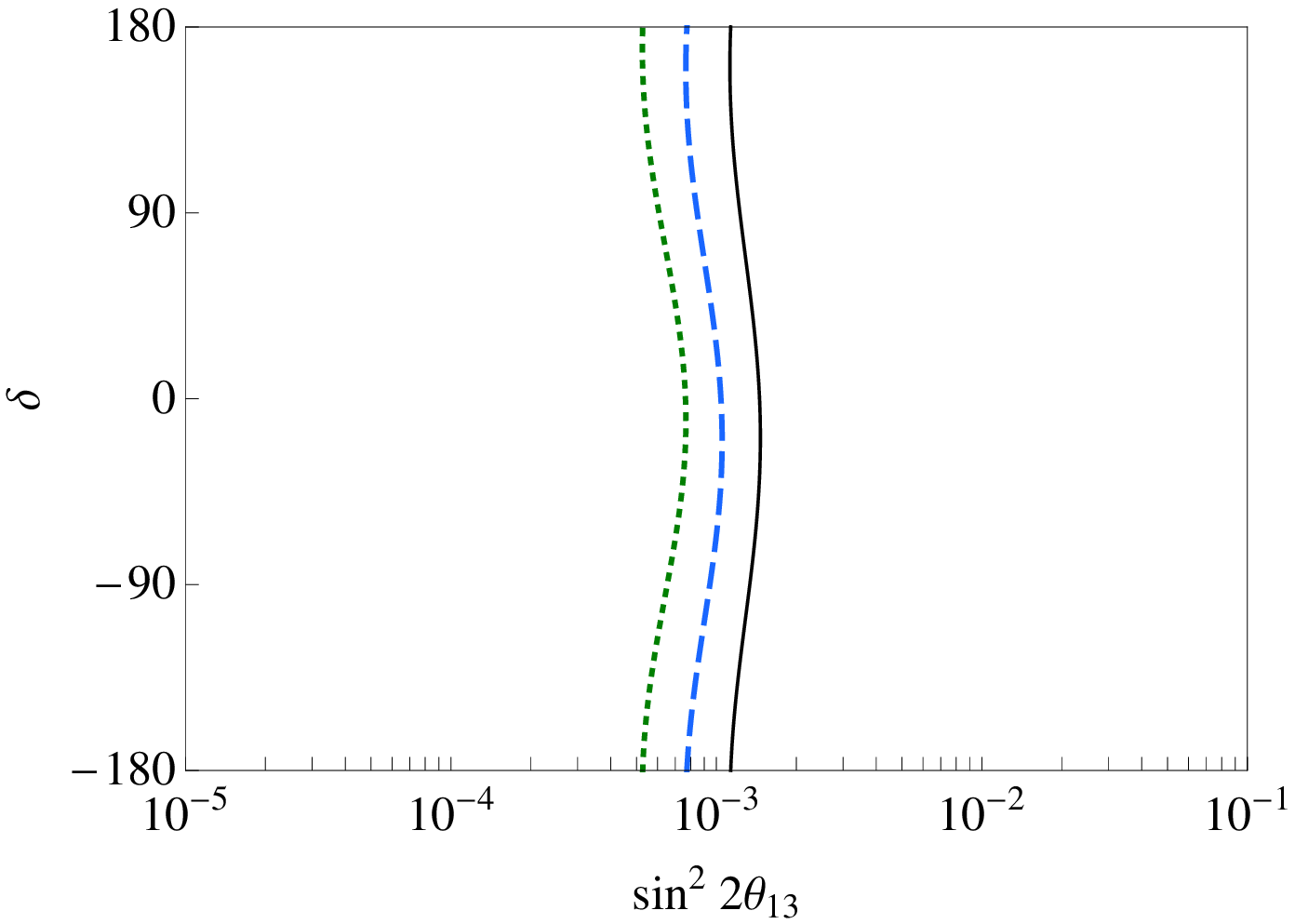}
\hglue 0.2cm
\includegraphics[width=0.46\textwidth,angle=0]{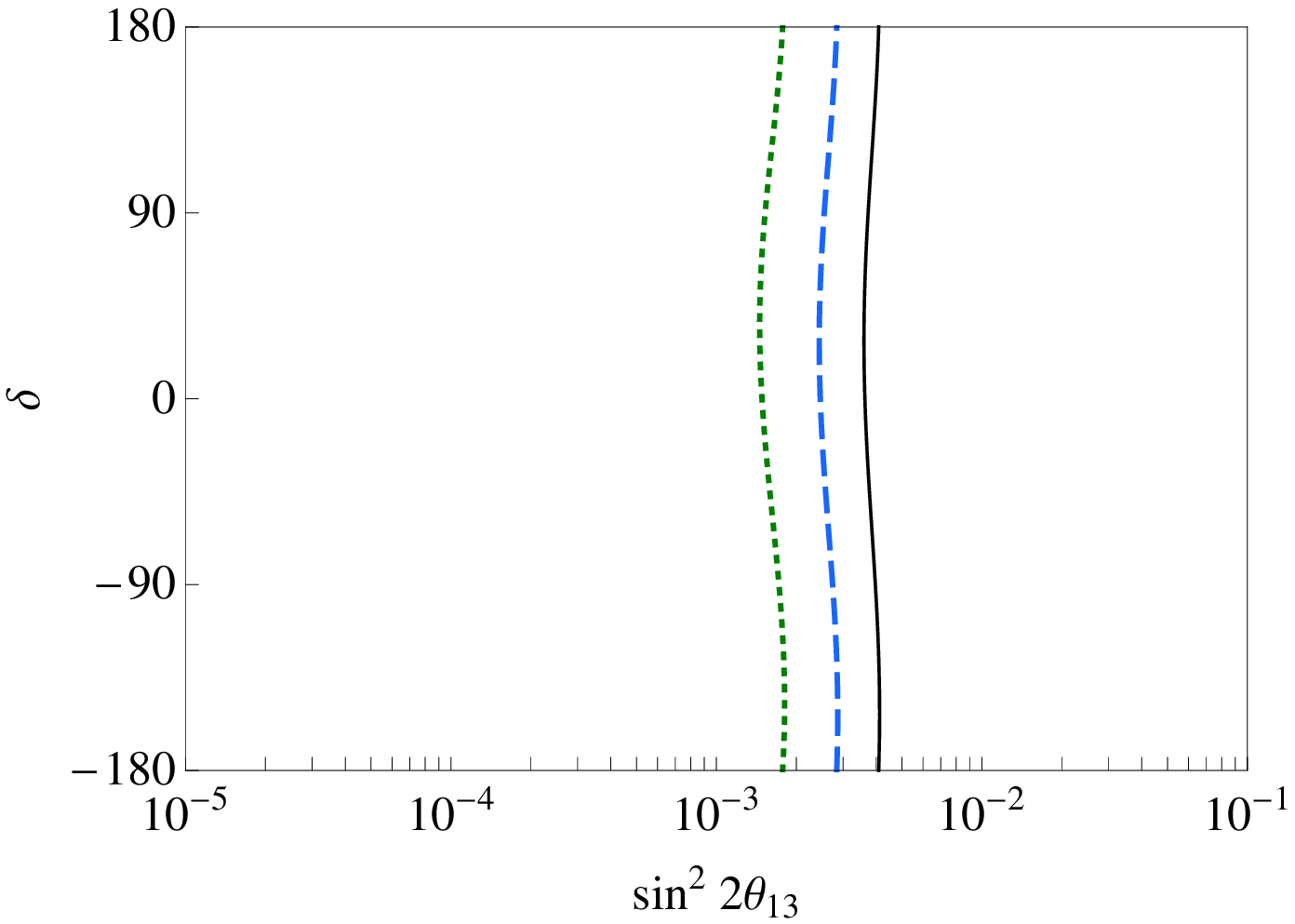}
\caption{\label{fig:sensring}
Comparison of the set-up proposed in this paper (black solid lines) with a set-up with 
longer decay rings (blue dashed lines) and longer decay plus improved  
detector characteristics (green dotted lines). Comparisons are 
shown for the three performance indicators and the layout of the 
panels are as for Fig. \ref{fig:sens}.}
\end{figure}

As stressed before, we have made a very conservative proposal for the 
two-baseline \bb set-up. 
In this subsection we study how stable the results presented here are to
modifications of the experimental set-up described. In particular, we focus on
two effects. 
The first 
is the gain in number of useful ion decays by increasing the length of 
the straight sections of the storage ring. 
The second is the uncertainty on the achievable low energy
threshold, efficiency and background at the iron detector. 
The sensitivity reach of 
our proposed set-up is shown in Fig. \ref{fig:sensring} by 
the black solid lines. We first probe the effect of increasing 
the number of useful ion decays by increasing the length of 
the straight sections of the storage ring for \br and \li ions.
This is shown by the blue dashed lines where we restore 
the straight sections to 2500 km. This increases the \br and 
\li flux at the far detector by 40\% compared to the 
black reference lines of our set-up.  
As it can be seen from the figure the impact
of increasing the flux at the far detector is mainly on the sensitivity to
the mass hierarchy (that becomes $\stch \leq 8 \times 10^{-4} (3 \times 10^{-3})$ for normal and inverted hierarchy, respectively), but is still mild even for that observable. 
Smaller and more feasible designs of the decay rings are therefore 
possible without affecting significantly the physics reach of the proposed facility.

The second effect concerns the detector specifications. 
For the reference set-up (black solid lines) we have assumed the same efficiencies and
backgrounds as a function of the neutrino energy as those derived for the MIND
detector when exposed to a Neutrino Factory beam in Ref.~\cite{Abe:2007bi}. As we
argued above, this is a conservative choice for the $\beta$-Beam, since this 
purer beam does not demand charge ID. Also, 
the spectrum is not as wide in energy
as that of the Neutrino Factory and hence the problems with neutral 
current backgrounds are also less severe. 
However, the task of the iron detector at the
long baseline is to determine the mass hierarchy and this will be achieved as
long as the efficiency at around $6-7$ GeV is high enough to observe the
matter resonance enhancement. The effect that a more optimistic assumption of
a lower energy threshold of $1.5$ GeV with a flat efficiency of $70\%$ and
background of $10^{-4}$ would imply for the different observables is shown in
Fig.~\ref{fig:sensring} by the green dotted lines. For these lines 
we also work with the longer decay ring with 40\% more fluxes at the far detector. 
As it can be seen from the
figure the gain is not very significant for any of the observables, this
confirms that the challenging efficient discrimination of the lowest energy
events mandatory for a Neutrino Factory, is not as critical for the set-up
proposed here.

%%%%%%%%%%%%%%%%%%%%%%%%%%%
\section{Conclusions}
%%%%%%%%%%%%%%%%%%%%%%%

We have presented a new $\beta$-Beam set-up that combines the strengths of the best set-ups in the literature trying to probe
with the same facility the key remaining unknown neutrino oscillation parameters: $\theta_{13}$, the existence of leptonic
CP violation and the neutrino mass ordering in the challenging regime of small $\theta_{13}$, trying 
to make the storage ring design more realistic than in previous studies.

The best CP discovery potentials can be achieved at low energies and short baselines that guarantee that matter 
effects are not strong enough to mimic true CP violation and spoil the measurement of $\delta$. 
On the other hand, the statistics at the detector grows with the $\gamma$ factor to which the ions are accelerated. We therefore follow Ref.~\cite{bc,bc2} for a compromise, choosing the highest $\gamma$ accessible at the SPS+ but exploiting the decay of the ions
with smallest end-point energy, that is, $^6$He and $^{18}$Ne. This guarantees good statistics at the detector, since the flux and
cross sections grow with $\gamma$, while maintaining a relatively low energy around $1$~GeV, which allows to consider the detection via
a Mton class water \chr detector. Furthermore, the oscillation baseline can be kept short, matching the CERN to Canfranc baseline of $650$~km, 
so as to further increase the statistics and to avoid strong matter effects that could spoil the CP discovery potential. 

On the other hand, the small matter effects preferred for the $\delta$ measurement
strongly limit the sensitivity of this set-up to the mass hierarchy. We then
consider the opposite regime, proposed in \cite{betaino1},  where these effects are strongest in order to improve this situation, that is, the resonant enhancement due to the matter interactions. The resonance occurs for energies around $6$~GeV, these energies can only be attained at a $\beta$-Beam combining
high $\gamma$ with ions with large end-point energy, like $^8$Li and
$^8$B. The enhancement will only take place in the (anti)neutrino channel if
the hierarchy is normal (inverted) therefore providing a very clean probe of
the mass ordering. Since the neutrino energies are in the multi-GeV regime,
one could use a 50 kton magnetized iron calorimeter as the far detector option.  
Moreover, a long baseline is required so that the density encountered by the neutrino beam is 
high enough. If the baseline is chosen to be close to the magic baseline, where all the dependence in $\delta$ is lost, the possible intrinsic degeneracies between $\theta_{13}$ and $\delta$ are also solved, thus increasing the synergy between the two baselines further. We then believe that the combination of the four ions and two baselines will provide the best $\beta$-Beam sensitivity to the remaining unknown neutrino oscillation parameters. 

While two-baseline \bb set-ups have been proposed and studied before, our
proposal is unique. We propose two different racetrack
geometry decay rings -- one for storing the \neon and \he ions, and another 
for storing the \br and \li ions. For magnetic field strength of 8.3 T, 
the storage ring for \neon and \he with $\gamma=350$, has straight
sections of 2500 km (as in the original proposal by Piero Zucchelli)
and hence a livetime fraction of 0.28. 
Since the neutrino beams generated from $^{18}$Ne
and $^6$He 
are sent over a baseline $L=650$ km to Canfranc, the maximum depth at the far
end of the storage ring has to be $d=197$ m only. The \br and \li beam, 
on the other hand, has to be sent over a baseline $L=7000$ km to INO, and hence its storage ring 
requires an inclination of $\vartheta=34.5^\circ$. For \li ions boosted at $\gamma = 350$, 
using the same ring as for the short baseline beam, this would require a maximum depth
$d=2132$ m at the far end of the storage ring. 
In order to alleviate the problem of the large $d$ needed for the beam going
to the magic baseline, one necessarily has to reduce the size of the straight
sections of the ring. This, however, would reduce the livetime and hence the 
number of muon events for the \br- and \li-generated neutrino beams. 
In order to compensate for this loss in the number of events, 
we propose to increase the $\gamma$ for the \br and \li ions. 
We point out that the number of events increase by 40\% with a small 10\% increase in the 
Lorentz boost from $\gamma=350$ to 390 for the \li ions. 
Therefore, we take $\gamma=390$ and $656$ for \li and \br ions respectively, as these are the 
limiting boost factors possible with the upgrades forecast for the SPS at CERN. 
This allows us to reduce the straight sections of the decay ring without significantly 
reducing the number of events at the detector, and 
hence the sensitivity of the experiment to oscillation parameters. 
Therefore, for a magnetic field of 8.3 T, one could have a decay ring with 
maximum depth at the far end of $d=1282$ m. Such a decay 
ring would give a livetime fraction of 0.17. 
We point out that magnetic fields as large at 15 T are under discussion for further
LHC upgrades.  With these larger magnets, 
one could design more compact decay rings with $d$ up to 1.8 times smaller 
for the ring for the magic baseline. For the ring for $L=650$ km, one 
could use the larger magnets to increase the livetime of the beam from {\em l} = 0.28 to {\em l} = 0.35. 

Even though the storage ring design proposed in this paper is more realistic than in former studies, it is still quite challenging. However, these are the kind of aggressive proposals being discussed for the next-to-next generation of facilities in order to probe CP violation and the mass hierarchy and to hunt for $\theta_{13}$ if it turns out to be beyond the sensitivity of the next generation of reactor and accelerator experiments. Indeed, the sensitivity gain that such a facility would provide compared to the combination of all the forthcoming reactor and accelerator experiments is remarkable. As can be seen from Fig.~1 of Ref.~\cite{Huber:2009cw} all the forthcoming facilities combined will be sensitive to $\theta_{13}$ down to
$\sin^2 2 \theta_{13} > 3-6 \times 10^{-3}$ at a $90 \%$ CL., the facility presented here would improve that sensitivity by one order of magnitude and with a $3 \sigma$ significance, see Fig.~\ref{fig:sens}. Even more striking is the gain in the ability to probe CP violation and the mass hierarchy. The discovery potential of CP violation of the forthcoming facilities is very limited, covering just a $20 \%$ of the $\delta$ parameter space and only if $\sin^2 2 \theta_{13} > 0.02$ at the $90 \%$ CL. Conversely, the setup proposed here would cover a $80 \%$ of the values of $\delta$ down to 
$\sin^2 2 \theta_{13} > 10^{-3}$ with still some sensitivity down to $\sin^2 2 \theta_{13} > 10^{-4}$ at $3 \sigma$.
As for the mass hierarchy, the combination of all the next generation experiments would grant a detection for less than $40 \%$ of the values of $\delta$ and $\sin^2 2 \theta_{13} > 0.04$ at a $90 \%$ CL., while the two-baseline $\beta$-Beam can go down to $\sin^2 2 \theta_{13} > 10^{-3}$ ($\sin^2 2 \theta_{13} > 3 \times 10^{-3}$) for normal (inverted) hierarchy at $3 \sigma$, regardless of the value of $\delta$.

\begin{figure}[t]
\includegraphics[width=0.46\textwidth,angle=0]{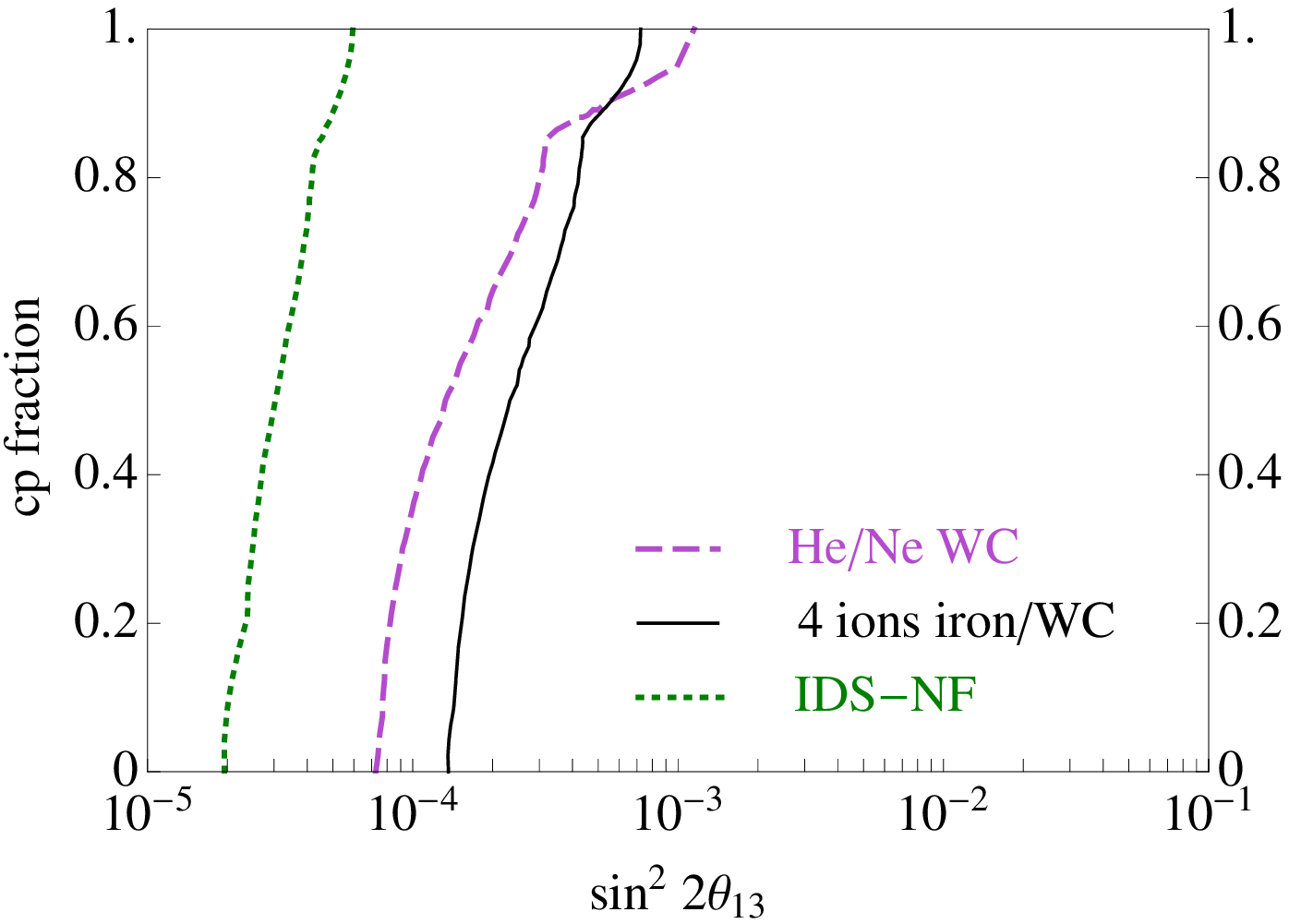}
\hglue 0.2cm
\includegraphics[width=0.46\textwidth,angle=0]{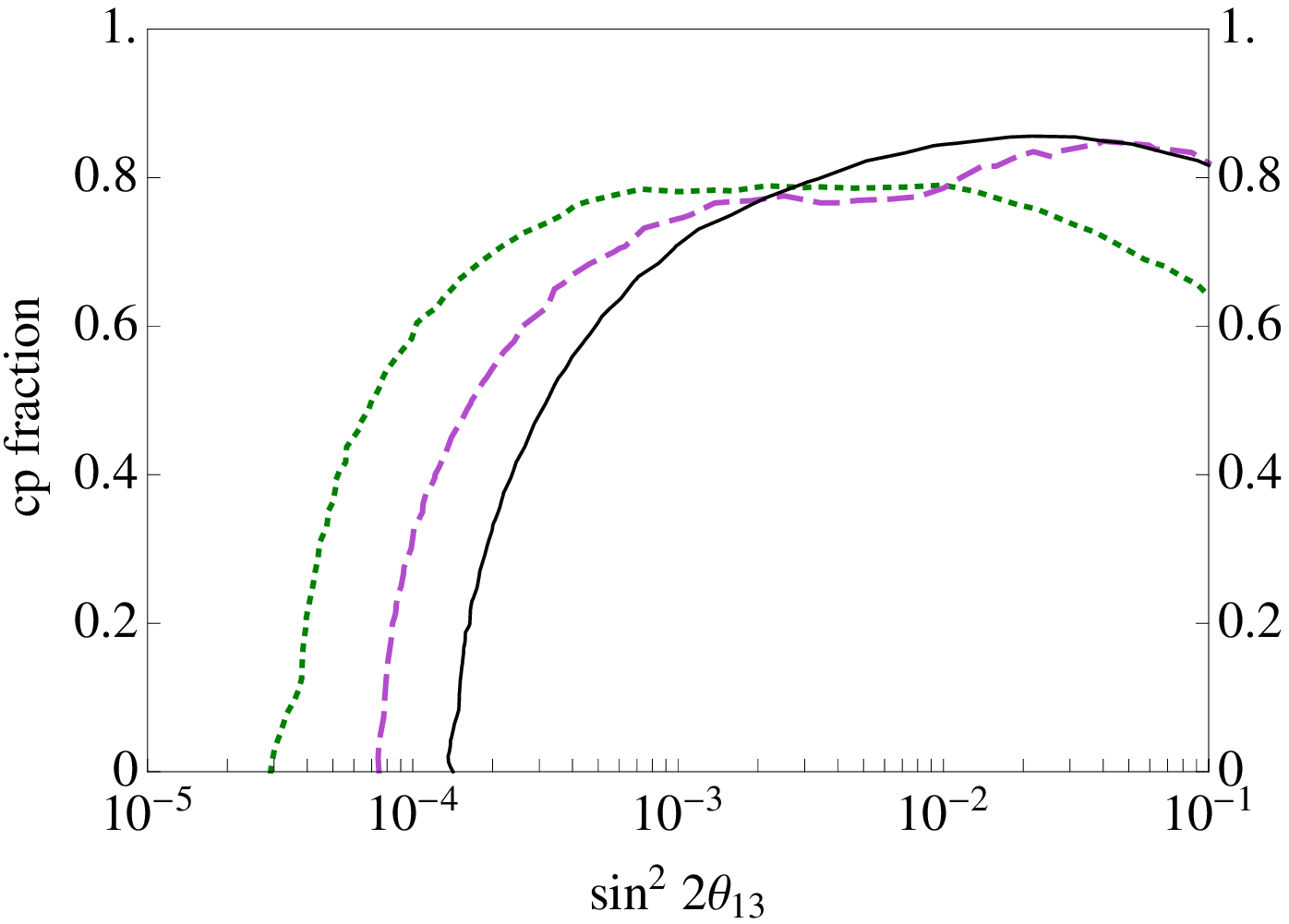}
\vglue 0.2cm
\includegraphics[width=0.46\textwidth,angle=0]{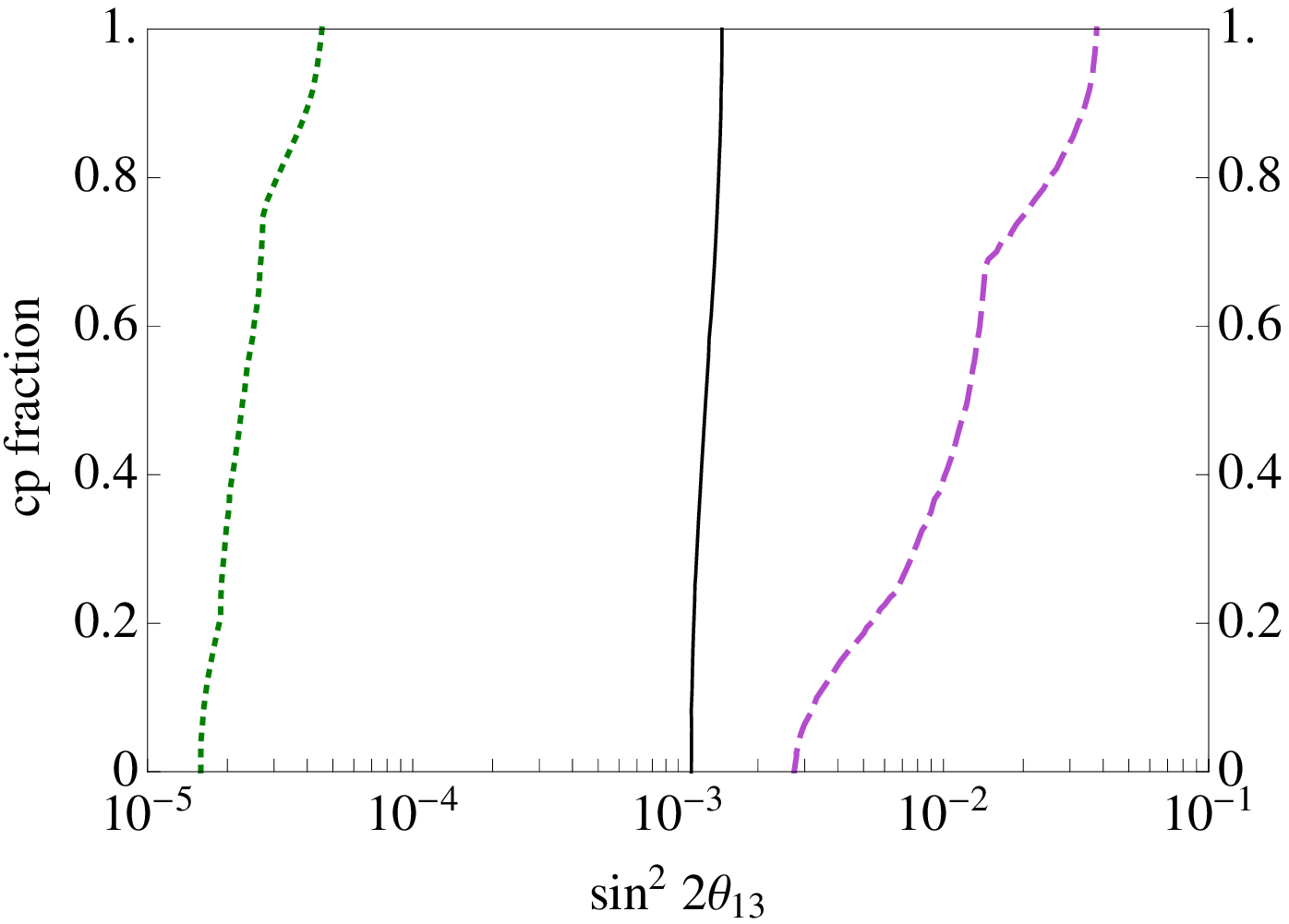}
\hglue 0.2cm
\includegraphics[width=0.46\textwidth,angle=0]{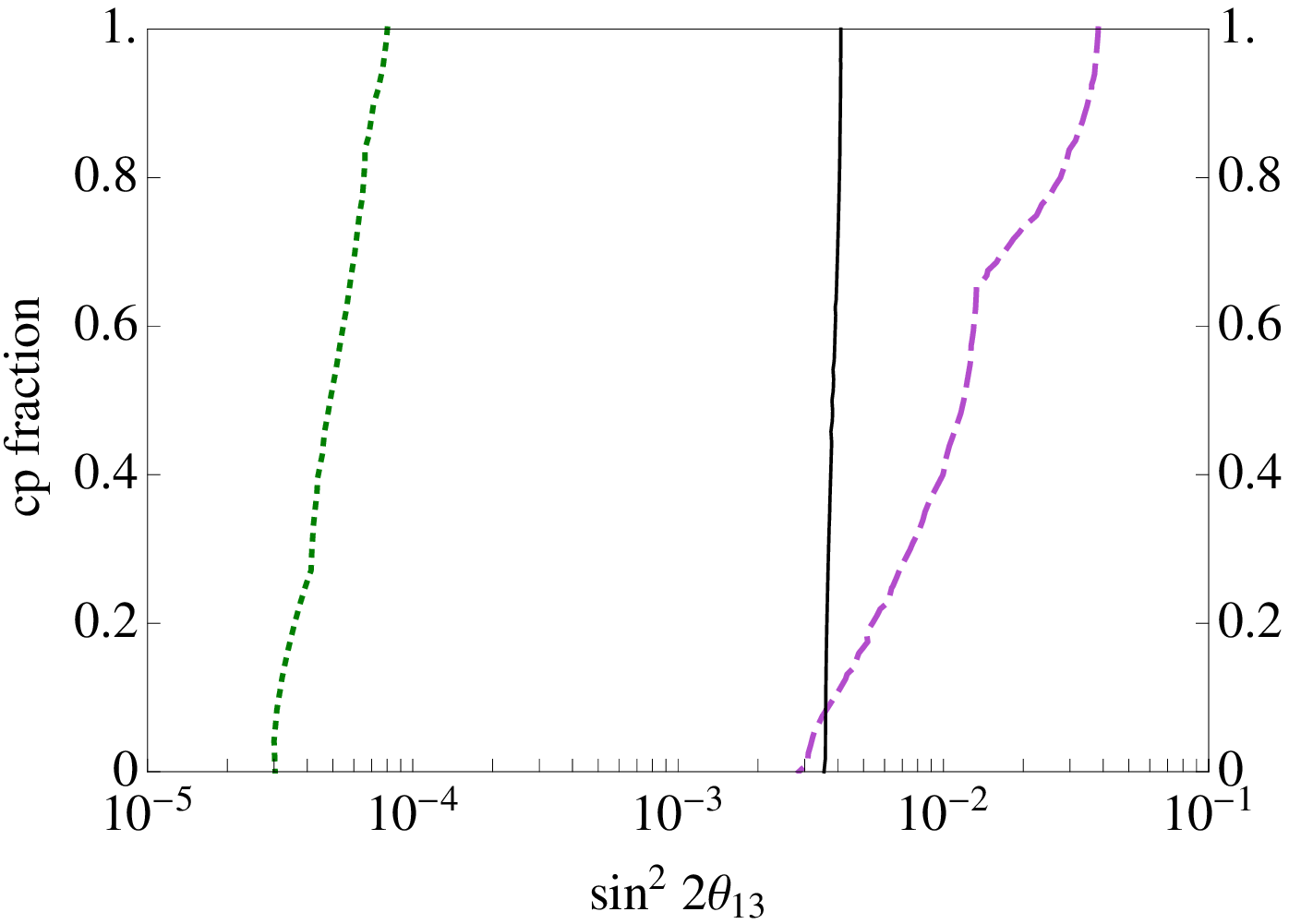}
\caption{\label{fig:senscpf}
Comparison of our proposed set-up (black solid lines) with 
the IDS Neutrino Factory baseline design (green dotted lines) and 
the high $\gamma$ \bb set-up from \cite{bc,bc2}.
The upper left hand panel 
shows $\stch$ discovery reach, the upper right hand 
panels shows the CP violation reach, while the lower 
panels show the mass hierarchy discovery reach. The 
lower left hand panel is for normal hierarchy as true 
while the lower right hand panel shows the corresponding 
reach when inverted hierarchy is true. }
\end{figure}

It is important to study how the sensitivities of the proposed set-up compare with other facilities of the next-to-next generation proposals. We compare the performance of our two-baseline $\beta$-Beam with the
other two facilities typically considered for the small $\theta_{13}$ regime: the IDS Neutrino Factory baseline design \cite{Bandyopadhyay:2007kx} and the high $\gamma$ $\beta$-Beam based on $^6$He and $^{18}$Ne of Ref.~\cite{bc,bc2}. For this comparison we present in Fig.~\ref{fig:senscpf} the same observables as in the previous figures but as a function of the fraction of the values of $\delta$ for which they can be discovered instead of the true values of $\delta$. This translates into a loss of information about the specific values of $\delta$ for which sensitivity is achieved but allows a better comparison of the relative performance of the different facilities.

From Fig.~\ref{fig:senscpf} it is clear that the facility with sensitivity to the different observables down to smallest values of $\stch$ is the Neutrino Factory. This can be understood from the very large fluxes assumed for the IDS baseline as compared to the ones assumed here for the $\beta$-Beam set-ups: $10^{21}$ useful muon decays per year to be compared to the $3 \times 10^{18}$ assumed for the $\beta$-Beams. This translates into much higher statistics that provide sensitivities to smaller values of $\theta_{13}$. On the other hand, the high energy of the Neutrino Factory beams implies a very small value of $L/E_\nu$. This translates in a stronger suppression of the CP violating term of the oscillation probability with respect to the one suppressed by two powers of $\theta_{13}$ for large values of this parameter. Therefore, the CP discovery potential of $\beta$-Beams outperforms that of the Neutrino Factory in Fig.~\ref{fig:senscpf} when $\sin^2 2 \theta_{13} > 10^{-3
 }$. 
 Since this large values of $\sin^2 2 \theta_{13}$ also guarantee a discovery of the mass hierarchy and $\stch$ regardless of the value of $\delta$, this makes $\beta$-Beams the better option when $\sin^2 2 \theta_{13} > 10^{-3}$. Furthermore, even if the statistics in the near $\beta$-Beam detector is reduced by half in the present set-up compared to the one in Ref.~\cite{bc,bc2} in order to illuminate the second detector, the CP-discovery potential for $\sin^2 2 \theta_{13} > 10^{-3}$ is better in the two-baseline set-up due to the lifting of the degeneracies
that can mimic CP-conservation when combining the information from the two detectors.

While the presently assumed $\beta$-Beam fluxes cannot compete with the expectations from a Neutrino Factory 
and cannot probe values of $\theta_{13}$ much smaller than $\stch \sim 10^{-4}$,
we find that $\beta$-Beam set-ups are better optimized for regions with $\sin^2 2 \theta_{13} > 10^{-3}$, providing
sensitivity to the different observables in larger fractions of the parameter space. In particular, 
we believe that the combination of ions and baselines proposed here represents an optimal \bb set-up,
that takes advantage of the properties of the different achievable beams, with very good sensitivity to 
all of the three observables considered, $\stch$, $\delta$ and the mass hierarchy 
(contrary to other \bb options, that are optimized for only one of them). 

%%%%%%%%%%%%%%%%%%%%%%%%%%%%%%%

\vglue 0.8cm
\noindent
{\Large{\bf Acknowledgments}}\vglue 0.3cm
\noindent
The authors wish to acknowledge S. Agarwalla for his contribution during early stages of this work. 
SC wishes to thank A. Raychaudhuri for discussions. 
SC and EFM would like to thank the Nordita program, 
``Astroparticle Physics - A Pathfinder to New Physics'' during which part of this work was done. 
AD thanks the IFIC at Valencia where part of this work was completed.
SC acknowledges support from the Neutrino Project under the XI Plan of Harish-Chandra Research Institute.
PC acknowledges financial support from the Comunidad Aut\'onoma de Madrid and from the 
Ministerio de Educaci\'on y Ciencia of Spain through project FPA2006-05423.
EFM acknowledges support by the DFG cluster of excellence ``Origin and Structure of the
Universe''. 
PC, AD and EFM also acknowledge support from the European
Community under the European Commission Framework Programme 7, Design
Study: EUROnu, Project Number 212372 (the EC is not liable for any use that may be made 
of the information contained herein). 
PC and AD eventually acknowledge support from the Spanish Government under
the Consolider-Ingenio 2010 programme: CUP, "Canfranc Underground Physics", Project Number CSD00C-08-44022.

%%%%%%%%%%%%%%%%%%%%%%%%%%%%%%%%%%%%%%%%%%%%%%%%%%%%%%%%%


\begin{thebibliography}{99}
\bibitem{solar}
B.~T.~Cleveland {\it et al.},
Astrophys.\ J.\  {\bf 496}, 505 (1998);
%%CITATION = ASJOA,496,505;%%
%
J.~N.~Abdurashitov {\it et al.}  [SAGE Collaboration],
J.\ Exp.\ Theor.\ Phys.\  {\bf 95}, 181 (2002)
[Zh.\ Eksp.\ Teor.\ Fiz.\  {\bf 122}, 211 (2002)];
%%CITATION = ZETFA,122,211;%%
%
W.~Hampel {\it et al.}  [GALLEX Collaboration],
Phys.\ Lett.\ B {\bf 447}, 127 (1999); 
%%CITATION = PHLTA,B447,127;%%
%
S.~Fukuda {\it et al.}  [Super-Kamiokande Collaboration],
Phys.\ Lett.\ B {\bf 539}, 179 (2002);
%%CITATION = PHLTA,B539,179;%%
%
Q.~R.~Ahmad {\it et al.}  [SNO Collaboration],
Phys.\ Rev.\ Lett.\  {\bf 89}, 011301 (2002);
%%CITATION = PRLTA,89,011301;%%
%
B.~Aharmim {\it et al.}  [SNO Collaboration],
Phys.\ Rev.\ C {\bf 72}, 055502 (2005);
%%CITATION = PHRVA,C72,055502;%%
%
B.~Aharmim {\it et al.}  [SNO Collaboration],
Phys.\ Rev.\ Lett.\  {\bf 101}, 111301 (2008);
%%CITATION = PRLTA,101,111301;%%
%
C.~Arpesella {\it et al.}
[Borexino~Collaboration],
Phys.\ Lett.\  B {\bf 658}, 101 (2008).
%%CITATION = PHLTA,B658,101;%%
\bibitem{atm}
  Y.~Ashie {\it et al.}  [Super-Kamiokande Collaboration],
%   ``A measurement of atmospheric neutrino oscillation parameters by
  %Super-Kamiokande I,''
  Phys.\ Rev.\ D {\bf 71}, 112005 (2005).
%  [arXiv:hep-ex/0501064].
%%CITATION = PHRVA,D71,112005;%%
\bibitem{kl}
  S.~Abe {\it et al.}  [KamLAND Collaboration],
  %``Precision Measurement of Neutrino Oscillation Parameters with KamLAND,''
  Phys.\ Rev.\ Lett.\  {\bf 100}, 221803 (2008).
%  [arXiv:0801.4589 [hep-ex]].
  %%CITATION = PRLTA,100,221803;%%
\bibitem{k2k}
E.~Aliu {\it et al.}  [K2K Collaboration],
  %``Evidence for muon neutrino oscillation in an accelerator-based
  %experiment,''
  Phys.\ Rev.\ Lett.\  {\bf 94}, 081802 (2005). 
\bibitem{minos}
  P.~Adamson {\it et al.}  [MINOS Collaboration],
  %``Measurement of Neutrino Oscillations with the MINOS Detectors in the NuMI
  %Beam,''
  Phys.\ Rev.\ Lett.\  {\bf 101}, 131802 (2008).
%  [arXiv:0806.2237 [hep-ex]].
  %%CITATION = PRLTA,101,131802;%%
\bibitem{limits}
  G.~L.~Fogli {\it et al.},
  %``Observables sensitive to absolute neutrino masses (Addendum),''
  Phys.\ Rev.\  D {\bf 78}, 033010 (2008)
  [arXiv:0805.2517 [hep-ph]];
  %%CITATION = PHRVA,D78,033010;%%
%
  T.~Schwetz, M.~Tortola and J.~W.~F.~Valle,
  %``Three-flavour neutrino oscillation update,''
  New J.\ Phys.\  {\bf 10}, 113011 (2008)
  [arXiv:0808.2016 [hep-ph]];
  %%CITATION = NJOPF,10,113011;%%
%
  A.~Bandyopadhyay, S.~Choubey, S.~Goswami, S.~T.~Petcov and D.~P.~Roy,
  %``Neutrino Oscillation Parameters After High Statistics KamLAND Results,''
  arXiv:0804.4857 [hep-ph].
  %%CITATION = ARXIV:0804.4857;%%
\bibitem{chooz}
M.~Apollonio {\it et al.},
%``Search for neutrino oscillations on a long base-line at the CHOOZ  nuclear
%power station,''
Eur.\ Phys.\ J.\ C {\bf 27}, 331 (2003).
%%CITATION = EPHJA,C27,331;%%
\bibitem{th13hint}
  G.~L.~Fogli, E.~Lisi, A.~Marrone, A.~Palazzo and A.~M.~Rotunno,
  %``Hints of theta_13>0 from global neutrino data analysis,''
  Phys.\ Rev.\ Lett.\  {\bf 101}, 141801 (2008)
  [arXiv:0806.2649 [hep-ph]];
  %%CITATION = PRLTA,101,141801;%%
%
  M.~Maltoni and T.~Schwetz,
  %``Three-flavour neutrino oscillation update and comments on possible hints
  %for a non-zero theta_{13},''
  arXiv:0812.3161 [hep-ph].
  %%CITATION = ARXIV:0812.3161;%%
\bibitem{th13hint2}
  A.~B.~Balantekin and D.~Yilmaz,
  %``Contrasting solar and reactor neutrinos with a non-zero value of theta13,''
  J.\ Phys.\ G {\bf 35}, 075007 (2008)
  [arXiv:0804.3345 [hep-ph]].
  %%CITATION = JPHGB,G35,075007;%%
\bibitem{chooz2}
  F.~Ardellier {\it et al.},
  %``Letter of intent for double-CHOOZ: A search for the mixing angle
  %theta(13),''
  arXiv:hep-ex/0405032;
  %%CITATION = HEP-EX/0405032;%%
%
  F.~Ardellier {\it et al.}  [Double Chooz Collaboration],
  %``Double Chooz: A search for the neutrino mixing angle theta(13),''
  arXiv:hep-ex/0606025.
  %%CITATION = HEP-EX/0606025;%%
\bibitem{dayabay}
  X.~Guo {\it et al.}  [Daya Bay Collaboration],
  %``A precision measurement of the neutrino mixing angle theta(13) using
  %reactor antineutrinos at Daya Bay,''
  arXiv:hep-ex/0701029.
  %%CITATION = HEP-EX/0701029;%%
\bibitem{reno}Information about the experiment available
at {\it http://neutrino.snu.ac.kr/RENO}
\bibitem{angra}
  J.~C.~Anjos {\it et al.},
  %``Angra neutrino project: Status and plans,''
  Nucl.\ Phys.\ Proc.\ Suppl.\  {\bf 155}, 231 (2006).
%  [arXiv:hep-ex/0511059].
  %%CITATION = NUPHZ,155,231;%%
\bibitem{t2k}
  Y.~Itow {\it et al.},
  %``The JHF-Kamioka neutrino project,''
  arXiv:hep-ex/0106019.
  %%CITATION = HEP-EX 0106019;%%
\bibitem{nova}
  D.~S.~Ayres {\it et al.}  [NOvA Collaboration],
  %``NOvA proposal to build a 30-kiloton off-axis detector to study neutrino
  %oscillations in the Fermilab NuMI beamline,''
  arXiv:hep-ex/0503053.
  %%CITATION = HEP-EX 0503053;%%
%\cite{Huber:2009cw}
\bibitem{Huber:2009cw}
  P.~Huber, M.~Lindner, T.~Schwetz and W.~Winter,
  %``First hint for CP violation in neutrino oscillations from upcoming
  %superbeam and reactor experiments,''
  arXiv:0907.1896 [hep-ph].
  %%CITATION = ARXIV:0907.1896;%%


\bibitem{geer}
  S.~Geer,
  %``Neutrino beams from muon storage rings: Characteristics and physics
  %potential,''
  Phys.\ Rev.\  D {\bf 57}, 6989 (1998)
  [Erratum-ibid.\  D {\bf 59}, 039903 (1999)].
%  [arXiv:hep-ph/9712290].
  %%CITATION = PHRVA,D57,6989;%%
\bibitem{zucc}
P.~Zucchelli,
%``A novel concept for a anti-nu/e / nu/e neutrino factory: The beta beam,''
Phys.\ Lett.\ B {\bf 532}, 166 (2002).
%%CITATION = PHLTA,B532,166;%%
\bibitem{lindroos}
 M.~Lindroos,
  %``The acceleration and storage of radioactive ions for a beta-beam
  %facility,''
  arXiv:physics/0312042;
  %%CITATION = PHYSICS/0312042;%%
%
  M.~Lindroos,
  %``The Technical Challenges Of Beta-Beams,''
  Nucl.\ Phys.\ Proc.\ Suppl.\  {\bf 155}, 48 (2006).
  %%CITATION = NUPHZ,155,48;%%
%
\bibitem{betabeampage}
http://beta-beam.web.cern.ch/beta\%2Dbeam/
%
\bibitem{golden}
  A.~Cervera, A.~Donini, M.~B.~Gavela, J.~J.~G\'{o}mez-Cadenas, P.~Hernandez, O.~Mena and S.~Rigolin,
  %``Golden measurements at a neutrino factory,''
  Nucl.\ Phys.\ B {\bf 579}, 17 (2000)
  [Erratum-ibid.\ B {\bf 593}, 731 (2001)].
%  [arXiv:hep-ph/0002108].
  %%CITATION = HEP-PH 0002108;%%
%
\bibitem{intrinsic}
  J.~Burguet-Castell, 
M.~B.~Gavela, J.~J.~G\'{o}mez-Cadenas, P.~Hernandez and O.~Mena,
  %``On the measurement of leptonic CP violation,''
  Nucl.\ Phys.\ B {\bf 608}, 301 (2001).
%  [arXiv:hep-ph/0103258].
%%CITATION = NUPHA,B608,301;%%
\bibitem{minadeg}
  H.~Minakata and H.~Nunokawa,
  %``Exploring neutrino mixing with low energy superbeams,''
  JHEP {\bf 0110}, 001 (2001).
%  [arXiv:hep-ph/0108085].
%%CITATION = JHEPA,0110,001;%%
\bibitem{th23octant}
  G.~L.~Fogli and E.~Lisi,
%   ``Tests of three-flavor mixing in long-baseline neutrino oscillation
  %experiments,''
  Phys.\ Rev.\ D {\bf 54}, 3667 (1996).
%  [arXiv:hep-ph/9604415].
%%CITATION = PHRVA,D54,3667;%%
\bibitem{eight}
  V.~Barger, D.~Marfatia and K.~Whisnant,
%   ``Breaking eight-fold degeneracies in neutrino CP violation, mixing, and
  %mass hierarchy,''
  Phys.\ Rev.\ D {\bf 65}, 073023 (2002).
%  [arXiv:hep-ph/0112119].
%%CITATION = PHRVA,D65,073023;%%
%
\bibitem{diffLnE}
  H.~Minakata and H.~Nunokawa,
  %``How to measure CP violation in neutrino oscillation experiments?,''
  Phys.\ Lett.\  B {\bf 413}, 369 (1997);
%  [arXiv:hep-ph/9706281].
  %%CITATION = PHLTA,B413,369;%%
%
  V.~Barger, D.~Marfatia and K.~Whisnant,
  %``Off-axis beams and detector clusters: Resolving neutrino parameter
  %degeneracies,''
  Phys.\ Rev.\  D {\bf 66}, 053007 (2002);
%  [arXiv:hep-ph/0206038].
  %%CITATION = PHRVA,D66,053007;%%
%
  V.~Barger, D.~Marfatia and K.~Whisnant,
  %``How two neutrino superbeam experiments do better than one,''
  Phys.\ Lett.\  B {\bf 560}, 75 (2003);
%  [arXiv:hep-ph/0210428].
  %%CITATION = PHLTA,B560,75;%%
%
  O.~Mena and S.~J.~Parke,
  %``Untangling CP violation and the mass hierarchy in long baseline
  %experiments,''
  Phys.\ Rev.\  D {\bf 70}, 093011 (2004);
%  [arXiv:hep-ph/0408070].
  %%CITATION = PHRVA,D70,093011;%%
%
  O.~Mena Requejo, S.~Palomares-Ruiz and S.~Pascoli,
  %``Super-NOvA: A long-baseline neutrino experiment with two off-axis
  %detectors,''
  Phys.\ Rev.\  D {\bf 72}, 053002 (2005);
%  [arXiv:hep-ph/0504015].
  %%CITATION = PHRVA,D72,053002;%%
%
  M.~Ishitsuka, T.~Kajita, H.~Minakata and H.~Nunokawa,
  %``Resolving neutrino mass hierarchy and CP degeneracy by two identical
  %detectors with different baselines,''
  Phys.\ Rev.\  D {\bf 72}, 033003 (2005);
%  [arXiv:hep-ph/0504026].
  %%CITATION = PHRVA,D72,033003;%%
%
  K.~Hagiwara, N.~Okamura and K.~i.~Senda,
  %``Physics potential of T2KK: An extension of the T2K neutrino oscillation
  %experiment with a far detector in Korea,''
  Phys.\ Rev.\  D {\bf 76}, 093002 (2007).
%  [arXiv:hep-ph/0607255].
  %%CITATION = PHRVA,D76,093002;%%
%
\bibitem{t2ksimulation}
  P.~Huber, M.~Lindner and W.~Winter,
  %``Superbeams versus neutrino factories,''
  Nucl.\ Phys.\  B {\bf 645}, 3 (2002);
%  [arXiv:hep-ph/0204352].
  %%CITATION = NUPHA,B645,3;%%
%
  P.~Huber, M.~Lindner and W.~Winter,
  %``Synergies between the first-generation JHF-SK and NuMI superbeam
  %experiments,''
  Nucl.\ Phys.\  B {\bf 654}, 3 (2003).
%  [arXiv:hep-ph/0211300].
  %%CITATION = NUPHA,B654,3;%%

\bibitem{silver}
  A.~Donini, D.~Meloni and P.~Migliozzi,
  %``The silver channel at the neutrino factory,''
  Nucl.\ Phys.\  B {\bf 646}, 321 (2002);
%  [arXiv:hep-ph/0206034].
  %%CITATION = NUPHA,B646,321;%%
%
  D.~Autiero {\it et al.},
  %``The synergy of the golden and silver channels at the Neutrino Factory,''
  Eur.\ Phys.\ J.\  C {\bf 33}, 243 (2004).
%  [arXiv:hep-ph/0305185].
  %%CITATION = EPHJA,C33,243;%%

\bibitem{dissappear}
  A.~Donini, E.~Fernandez-Martinez and S.~Rigolin,
  %``Appearance and disappearance signals at a beta-beam and a super-beam
  %facility,''
  Phys.\ Lett.\  B {\bf 621}, 276 (2005);
%  [arXiv:hep-ph/0411402].
  %%CITATION = PHLTA,B621,276;%%
%
  A.~Donini, E.~Fernandez-Martinez, D.~Meloni and S.~Rigolin,
  %``nu/mu disappearance at the SPL, T2K-I, NOnuA and the neutrino factory,''
  Nucl.\ Phys.\  B {\bf 743}, 41 (2006).
%  [arXiv:hep-ph/0512038].
  %%CITATION = NUPHA,B743,41;%%
\bibitem{pee}
  S.~K.~Agarwalla, S.~Choubey, S.~Goswami and A.~Raychaudhuri,
  %``Neutrino parameters from matter effects in $P_{ee}$ at long baselines,''
  Phys.\ Rev.\  D {\bf 75}, 097302 (2007).
%  [arXiv:hep-ph/0611233].
  %%CITATION = PHRVA,D75,097302;%%

\bibitem{addatm}
  P.~Huber, M.~Maltoni and T.~Schwetz,
  %``Resolving parameter degeneracies in long-baseline experiments by
  %atmospheric neutrino data,''
  Phys.\ Rev.\  D {\bf 71}, 053006 (2005);
%  [arXiv:hep-ph/0501037].
  %%CITATION = PHRVA,D71,053006;%%
\bibitem{cernmemphys}
  J.~E.~Campagne, M.~Maltoni, M.~Mezzetto and T.~Schwetz,
  %``Physics potential of the CERN-MEMPHYS neutrino oscillation project,''
  JHEP {\bf 0704}, 003 (2007).
%  [arXiv:hep-ph/0603172].
  %%CITATION = JHEPA,0704,003;%%
\bibitem{addreact}
  P.~Huber, M.~Lindner, T.~Schwetz and W.~Winter,
  %``Reactor Neutrino Experiments Compared to Superbeams,''
  Nucl.\ Phys.\  B {\bf 665}, 487 (2003).
%  [arXiv:hep-ph/0303232].
  %%CITATION = NUPHA,B665,487;%%

\bibitem{magic}
  P.~Huber and W.~Winter,
  %``Neutrino factories and the 'magic' baseline,''
  Phys.\ Rev.\ D {\bf 68}, 037301 (2003).
%  [arXiv:hep-ph/0301257].
%%CITATION = PHRVA,D68,037301;%%
\bibitem{magic2}
  A.~Y.~Smirnov,
  %``Neutrino oscillations: what is magic about the magic baseline?,''
  arXiv:hep-ph/0610198.
  %%CITATION = HEP-PH 0610198;%%
\bibitem{petcov}
  M.~Freund, M.~Lindner, S.~T.~Petcov and A.~Romanino,
  %``Testing matter effects in very long baseline neutrino oscillation
  %experiments,''
  Nucl.\ Phys.\  B {\bf 578}, 27 (2000).
  %%CITATION = NUPHA,B578,27;%%

\bibitem{nufactoptim}
  P.~Huber, M.~Lindner, M.~Rolinec and W.~Winter,
  %``Optimization of a neutrino factory oscillation experiment,''
  Phys.\ Rev.\  D {\bf 74}, 073003 (2006)
  [arXiv:hep-ph/0606119]
  %%CITATION = PHRVA,D74,073003;%%
\bibitem{bboptim}
  S.~K.~Agarwalla, S.~Choubey, A.~Raychaudhuri and W.~Winter,
  %``Optimizing the greenfield Beta-beam,''
  JHEP {\bf 0806}, 090 (2008)
  [arXiv:0802.3621 [hep-ex]].
  %%CITATION = JHEPA,0806,090;%%

\bibitem{rubbia}
    C.~Rubbia, A.~Ferrari, Y.~Kadi and V.~Vlachoudis,
  %``Beam cooling with ionisation losses,''
  Nucl.\ Instrum.\ Meth.\ A {\bf 568}, 475 (2006);
%  [arXiv:hep-ph/0602032].
  %%CITATION = HEP-PH 0602032;%%
  C.~Rubbia,
%   ``Ionization cooled ultra pure beta-beams for long distance nu/e --> nu/mu
  %transitions, theta(13) phase and CP violation,''
  arXiv:hep-ph/0609235.
  %%CITATION = HEP-PH 0609235;%%
\bibitem{mori}
% Y.~Mori, Nucl.\ Instrum.\ Meth.\ A {\bf 562}, 591 (2006).
  Y.~Mori,
  %``Development of FFAG accelerators and their applications for
  %intense secondary particle production,''
  Nucl.\ Instrum.\ Meth.\  A {\bf 562}, 591 (2006).
  %%CITATION = NUIMA,A562,591;%%

\bibitem{betaino1}
  S.~K.~Agarwalla, S.~Choubey and A.~Raychaudhuri,
  %``Neutrino mass hierarchy and theta(13) with a magic baseline beta-beam
  %experiment,''
  Nucl.\ Phys.\  B {\bf 771}, 1 (2007)
  [arXiv:hep-ph/0610333].
  %%CITATION = NUPHA,B771,1;%%

\bibitem{betaino2}
  S.~K.~Agarwalla, S.~Choubey and A.~Raychaudhuri,
  %``Unraveling neutrino parameters with a magical beta-beam experiment at
  %INO,''
  Nucl.\ Phys.\  B {\bf 798}, 124 (2008)
  arXiv:0711.1459 [hep-ph].
  %%CITATION = ARXIV:0711.1459;%%

\bibitem{twobaseline1}
  P.~Coloma, A.~Donini, E.~Fernandez-Martinez and J.~Lopez-Pavon,
  %``$\theta_{13}$, $\delta$ and the neutrino mass hierarchy at a $\gamma=350$
  %double baseline Li/B $\beta$-Beam,''
  JHEP {\bf 0805}, 050 (2008)
  [arXiv:0712.0796 [hep-ph]].
  %%CITATION = JHEPA,0805,050;%%

\bibitem{twobaseline2}
  S.~K.~Agarwalla, S.~Choubey and A.~Raychaudhuri,
  %``Exceptional Sensitivity to Neutrino Parameters with a Two Baseline
  %Beta-Beam Set-up,''
  Nucl.\ Phys.\  B {\bf 805}, 305 (2008)
  [arXiv:0804.3007 [hep-ph]].
  %%CITATION = NUPHA,B805,305;%%
  
  \bibitem{bc2}
  J.~Burguet-Castell, D.~Casper, J.~J.~G\'{o}mez-Cadenas, P.~Hernandez and F.~Sanchez,
  %``Neutrino oscillation physics with a higher gamma beta-beam,''
  Nucl.\ Phys.\  B {\bf 695}, 217 (2004)
  [arXiv:hep-ph/0312068].
  %%CITATION = NUPHA,B695,217;%%

\bibitem{PAF}
  ``Proton Accelerator for the Future'' (PAF) inter-departmental working group webpage, http://pofpa.web.cern.ch/pofpa/

\bibitem{beta}
L. P. Ekstrom and R. B. Firestone, WWW Table of Radioactive Isotopes, \\
database version 2/28/99 from URL http://ie.lbl.gov/toi/

%\cite{Donini:2008zz}
\bibitem{Donini:2008zz}
  A.~Donini and M.~Lindroos,
  %``Optimisation of a beta beam,''
%\href{http://www.slac.stanford.edu/spires/find/hep/www?irn=8307784}{SPIRES entry}
{\it Presented at 10th International Workshop on Neutrino Factories,   Superbeams and Betabeams: Nufact08, Valencia, Spain, 30 Jun - 5 Jul 2008}

  %\cite{Freund:2001pn}
\bibitem{Freund:2001pn}
  M.~Freund,
  %``Analytic approximations for three neutrino oscillation parameters and
  %probabilities in matter,''
  Phys.\ Rev.\  D {\bf 64} (2001) 053003
  [arXiv:hep-ph/0103300].
  %%CITATION = PHRVA,D64,053003;%%

\bibitem{msw1}
  L.~Wolfenstein,
  Phys.\ Rev.\ D {\bf 17}, 2369 (1978);
%%CITATION = PHRVA,D17,2369;%%
\bibitem{msw2}
  S.~P.~Mikheev and A.~Y.~Smirnov,
  Sov.\ J.\ Nucl.\ Phys.\  {\bf 42}, 913 (1985)
  [Yad.\ Fiz.\  {\bf 42}, 1441 (1985)];
%%CITATION = YAFIA,42,1441;%%
%
  S.~P.~Mikheev and A.~Y.~Smirnov,
  Nuovo Cim.\ C {\bf 9}, 17 (1986).
%%CITATION = NUCIA,9C,17;%%
\bibitem{msw3}
  V.~D.~Barger, K.~Whisnant, S.~Pakvasa and R.~J.~N.~Phillips,
  %``Matter effects on three-neutrino oscillations,''
  Phys.\ Rev.\ D {\bf 22}, 2718 (1980).
%%CITATION = PHRVA,D22,2718;%%

\bibitem{prem}
  A.~M.~Dziewonski and D.~L.~Anderson,
  %``Preliminary Reference Earth Model,''
  Phys.\ Earth Planet.\ Interiors {\bf 25}, 297 (1981);
%%CITATION = PEPIA,25,297;%%
\\
S.~V.~Panasyuk, Reference Earth Model (REM) webpage,\\
 http://cfauves5.harvrd.edu/lana/rem/index.html.

  \bibitem{Zisman:2008zz}
  M.~S.~Zisman,
  %``International Scoping Study: Accelerator Working Group report,''
  J.\ Phys.\ Conf.\ Ser.\  {\bf 110} (2008) 112006.
  %%CITATION = 00462,110,112006;%%

\bibitem{Bandyopadhyay:2007kx}
  A.~Bandyopadhyay {\it et al.}  [ISS Physics Working Group],
  %``Physics at a future Neutrino Factory and super-beam facility,''
  arXiv:0710.4947 [hep-ph].
  %%CITATION = ARXIV:0710.4947;%%
  %\cite{Zisman:2008zz}

\bibitem{Geer:2007kn}
  S.~Geer, O.~Mena and S.~Pascoli,
  %``A Low energy neutrino factory for large $\theta_{13}$,''
  Phys.\ Rev.\  D {\bf 75} (2007) 093001
  [arXiv:hep-ph/0701258].
  %%CITATION = PHRVA,D75,093001;%%

%\cite{Bross:2007ts}
\bibitem{Bross:2007ts}
  A.~D.~Bross, M.~Ellis, S.~Geer, O.~Mena and S.~Pascoli,
  %``A Neutrino factory for both large and small $\theta_{13}$,''
  Phys.\ Rev.\  D {\bf 77} (2008) 093012
  [arXiv:0709.3889 [hep-ph]].
  %%CITATION = PHRVA,D77,093012;%%

\bibitem{ino}
  M.~S.~Athar {\it et al.}  [INO Collaboration],
%``India-based Neutrino Observatory: Project Report. Volume I,''
%INO-2006-01
%\href{http://www.slac.stanford.edu/spires/find/hep/www?r=ino-2006-01}{SPIRES entry}
 A Report of the INO Feasibility Study,\\
{http://www.imsc.res.in/~ino/OpenReports/INOReport.pdf}
%%CITATION = INO-2006-01;%%
%

\bibitem{betaoptim}
  P.~Huber, M.~Lindner, M.~Rolinec and W.~Winter,
  %``Physics and optimization of beta-beams: From low to very high gamma,''
  Phys.\ Rev.\  D {\bf 73}, 053002 (2006)
  [arXiv:hep-ph/0506237]
  %%CITATION = PHRVA,D73,053002;%%
  
  \bibitem{bc}
  J.~Burguet-Castell, D.~Casper, E.~Couce, J.~J.~G\'{o}mez-Cadenas and P.~Hernandez,
  %``Optimal beta-beam at the CERN-SPS,''
  Nucl.\ Phys.\  B {\bf 725}, 306 (2005)
  [arXiv:hep-ph/0503021].
  %%CITATION = NUPHA,B725,306;%%

\bibitem{oldpapers}
  M.~Mezzetto,
  %``Physics reach of the beta beam,''
  J.\ Phys.\ G {\bf 29}, 1771 (2003)
  [arXiv:hep-ex/0302007];
  %%CITATION = JPHGB,G29,1771;%%
%
  M.~Mezzetto,
  %``Beta beams,''
  Nucl.\ Phys.\ Proc.\ Suppl.\  {\bf 143}, 309 (2005)
  [arXiv:hep-ex/0410083];
  %%CITATION = NUPHZ,143,309;%%
%
  M.~Mezzetto,
  %``Physics potential of the gamma = 100,100 beta beam,''
  Nucl.\ Phys.\ Proc.\ Suppl.\  {\bf 155}, 214 (2006)
  [arXiv:hep-ex/0511005].
  %%CITATION = NUPHZ,155,214;%%
%
\bibitem{donini130}
  A.~Donini, E.~Fernandez-Martinez, P.~Migliozzi, S.~Rigolin and 
L.~Scotto Lavina,
  %``Study of the eightfold degeneracy with a standard beta-beam and a
  %super-beam facility,''
  Nucl.\ Phys.\  B {\bf 710}, 402 (2005)
  [arXiv:hep-ph/0406132].
  %%CITATION = NUPHA,B710,402;%%

\bibitem{doninialter}
  A.~Donini and E.~Fernandez-Martinez,
  %``Alternating ions in a beta-beam to solve degeneracies,''
  Phys.\ Lett.\ B {\bf 641}, 432 (2006)
  [arXiv:hep-ph/0603261].
%%CITATION = PHLTA,B641,432;%%
\bibitem{fnal}
  A.~Jansson, O.~Mena, S.~J.~Parke and N.~Saoulidou,
  %``Combining CPT-conjugate Neutrino channels at Fermilab,''
  Phys.\ Rev.\  D {\bf 78}, 053002 (2008)
  [arXiv:0711.1075 [hep-ph]].
  %%CITATION = PHRVA,D78,053002;%%
\bibitem{boulby}
  D.~Meloni, O.~Mena, C.~Orme, S.~Palomares-Ruiz and S.~Pascoli,
  %``An intermediate gamma beta-beam neutrino experiment with long baseline,''
  JHEP {\bf 0807}, 115 (2008)
  [arXiv:0802.0255 [hep-ph]].
  %%CITATION = JHEPA,0807,115;%%
  
  \bibitem{paper1}
  S.~K.~Agarwalla, A.~Raychaudhuri and A.~Samanta,
%   ``Exploration prospects of a long baseline beta beam neutrino experiment
  %with an iron calorimeter detector,''
  Phys.\ Lett.\ B {\bf 629}, 33 (2005)
  [arXiv:hep-ph/0505015].
  %%CITATION = HEP-PH 0505015;%%

\bibitem{doninibeta}
  A.~Donini, E.~Fernandez, P.~Migliozzi, S.~Rigolin, L.~Scotto Lavina, 
  T.~Tabarelli de Fatis and F.~Terranova,
  %``Perspectives for a neutrino program based on the upgrades of the CERN
  %accelerator complex,''
  arXiv:hep-ph/0511134;
%%CITATION = HEP-PH/0511134;%%
%
  A.~Donini, E.~Fernandez-Martinez, P.~Migliozzi, S.~Rigolin, 
  L.~Scotto Lavina, T.~Tabarelli de Fatis and F.~Terranova,
  %``A beta beam complex based on the machine upgrades of the LHC,''
  Eur.\ Phys.\ J.\  C {\bf 48}, 787 (2006).
%  [arXiv:hep-ph/0604229].
  %%CITATION = EPHJA,C48,787;%%

\bibitem{rparity}
  R.~Adhikari, S.~K.~Agarwalla and A.~Raychaudhuri,
%   ``Can R-parity violating supersymmetry be seen in long baseline beta-beam
  %experiments?,''
  Phys.\ Lett.\ B {\bf 642}, 111 (2006).
%  arXiv:hep-ph/0608034;
  %%CITATION = HEP-PH 0608034;%%
%
  S.~K.~Agarwalla, S.~Rakshit and A.~Raychaudhuri,
  %``Probing lepton number violating interactions with beta-beams,''
  Phys.\ Lett.\  B {\bf 647}, 380 (2007).
%  [arXiv:hep-ph/0609252].
  %%CITATION = PHLTA,B647,380;%%

%\cite{Abe:2007bi}
\bibitem{Abe:2007bi}
  T.~Abe {\it et al.}  [ISS Detector Working Group],
  %``Detectors and flux instrumentation for future neutrino facilities,''
  arXiv:0712.4129 [physics.ins-det].
  %%CITATION = ARXIV:0712.4129;%%

%\cite{deBellefon:2006vq}
\bibitem{Memphys}
  A.~de Bellefon {\it et al.},
  %``MEMPHYS: A large scale water Cerenkov detector at Frejus,''
  arXiv:hep-ex/0607026.
  %%CITATION = HEP-EX/0607026;%

  \bibitem{mind}
A. Cervera, Talk at NuFact07, Okayama University, Okayama, Japan, 
August 6-11, 2007, http://fphy.hep.okayama-u.ac.jp/nufact07/

  %\cite{Huber:2004ka}
\bibitem{globes1}
  P.~Huber, M.~Lindner and W.~Winter,
  %``Simulation of long-baseline neutrino oscillation experiments with
  %GLoBES,''
  Comput.\ Phys.\ Commun.\  {\bf 167} (2005) 195
  [arXiv:hep-ph/0407333].
  %%CITATION = CPHCB,167,195;%%
 
%\cite{Huber:2007ji}
\bibitem{globes2}
  P.~Huber, J.~Kopp, M.~Lindner, M.~Rolinec and W.~Winter,
  %``New features in the simulation of neutrino oscillation experiments with
  %GLoBES 3.0,''
  Comput.\ Phys.\ Commun.\  {\bf 177} (2007) 432
  [arXiv:hep-ph/0701187].
  %%CITATION = CPHCB,177,432;%%
  
  %\cite{Huber:2002mx}
\bibitem{Huber:2002mx}
  P.~Huber, M.~Lindner and W.~Winter,
  %``Superbeams versus neutrino factories,''
  Nucl.\ Phys.\  B {\bf 645} (2002) 3
  [arXiv:hep-ph/0204352].
  %%CITATION = NUPHA,B645,3;%% 

\end{thebibliography}
\end{document}